\documentclass[twocolumn,notitlepage,tightenlines,showpacs,preprintnumbers,amsmath,amssymb]{revtex4}

\usepackage{graphics}
\usepackage{graphicx,amsmath}
\usepackage{epsf}
\usepackage{epstopdf}
\usepackage{hyperref}
\usepackage[]{units}
\usepackage{ifpdf}

\begin{document}
\title{Optomechanical Multi-Mode Hamiltonian for Nanophotonic Waveguides}
\author{Hashem~Zoubi}
\email{hashem.zoubi@itp.uni-hannover.de}
\author{Klemens~Hammerer}
\affiliation{Institute for Theoretical Physics, Institute for Gravitational
  Physics (Albert Einstein Institute), Leibniz University Hannover,
  Callinstrasse 38, 30167 Hannover, Germany}
\date{03 August 2016}

\begin{abstract}
We develop a systematic method for deriving a quantum optical multi-mode
Hamiltonian for the interaction of photons and phonons in nanophotonic
dielectric materials by applying perturbation theory to the electromagnetic
Hamiltonian. The Hamiltonian covers radiation pressure and electrostrictive
interactions on equal footing. As a paradigmatic example, we apply our method
to a cylindrical nanoscale waveguide, and derive a Hamiltonian description of
Brillouin quantum optomechanics. We show analytically that in nanoscale waveguides radiation pressure dominates over electrostriction, in agreement with recent experiments. The calculated photon-phonon coupling parameters are used to infer gain parameters of Stokes Brillouin scattering in good agreement with experimental observations.
\end{abstract}



\maketitle

\section{Introduction}

Quantum optomechanics is the study of phenomena originating from the mutual interaction between electromagnetic radiation and mechanical motion \cite{Aspelmeyer2010,Aspelmeyer2014,Aspelmeyer2014a,Bowen2016}. In cavity optomechanics both the electromagnetic field and the mechanical vibrations are effectively restricted to single modes, and the strong coupling among phonons and photons achievable there enabled the demonstration of various quantum mechanical effects during the last years \cite{Hammerer2014}. In the domain of optical frequencies strongest optomechanical coupling has been obtained in optomechanical crystals \cite{Eichenfield2009,Safavi-Naeini2014} where nanostructuring of dielectric materials is exploited to generate phonon and photon modes with strong spatial localization in order to enhance the light-matter interactions.

Very recently, experimental progress with nanophotonic waveguides supporting
long-lived, high-frequency phonon modes evidenced that quantum optomechanical
effects may become accessible also within a multi-mode version of
optomechanics involving continua of propagating phonon and photon modes
\cite{VanLaer2015,Rakish2012,Shin2013,VanLaer2015a,VanLaer2015b,Kittlaus2015}. At
the classical level, the Brillouin physics of interacting photons and phonons in waveguides has been studied extensively and led to the demonstration of a wide scope of nonlinear optical phenomena, see \cite{Thevenaz2008,Kobyakov2010,Eggleton2013,Bahl2014} for reviews. So far, the dominant mechanism for optomechanical coupling in waveguides has been electrostriction, that is the modulation of the index of refraction of the bulk dielectric material associated with its acoustic vibrations causing scattering of photons on these periodic index modulations. The recent experiments with nanophotonic waveguides entered a new regime where radiation pressure effects due to vibrational surface deformations start to dominate over electrostriction which may result in vastly enhanced photon-phonon coupling. At the same time, these devices can maintain large quality factors for GHz  mechanical modes extending over long cm-scale nanowires providing large optomechanical interactions over significant 
 time and length scales. Overall, these developments indicate Brillouin quantum optomechanics as a promising route towards integrable, broad-band platforms supporting strongly interacting fields of phonons and photons.

On the theoretical side, the description of Brillouin optomechanics in classical terms is well established \cite{Fabelinskii1968,Boyd2008,Rakish2012,Wolff2015}. A corresponding quantum mechanical description of electrostrictive coupling in terms of a multi-mode Hamiltonian has recently been derived by Agrawal et al.~\cite{Agarwal2013a} in the context of multimode phonon cooling. In cavity quantum optomechanics involving single phonon and photon modes both effects are commonly taken into account \cite{Eichenfield2009,Safavi-Naeini2014} based on a formula by Johnson et al. \cite{Johnson2002}. Progress towards extending the cavity optomechanical description to the regime of Brillouin optomechanics has been achieved by Van Laer et
al. \cite{VanLaer2015c} by relating Brillouin gain parameters to the optomechanical single-photon coupling strengths. Sipe et al.~\cite{Sipe2016} very recently provided a Hamiltonian treatment of stimulated Brillouin scattering in nanoscale integrated waveguides accounting for electrostriction and radiation pressure, following the method of \cite{Johnson2002,Wolff2015}. 

In the present work we aim to contribute to this development of a quantum
theory of Brillouin optomechanics in two respects: Firstly, we derive the
multi-mode Hamiltonian for Brillouin optomechanics by applying perturbation
theory directly to the field Hamiltonian for the case of an isotropic
dielectric material which includes on equal footing electrostriction and radiation
pressure mechanisms. Our derivation reproduces the results of Sipe et
al.~\cite{Sipe2016} but avoids the rather technical smoothing procedures
introduced by Johnson et al.~\cite{Johnson2002}, employed also in
\cite{Sipe2016}, in order to deal with discontinuities at surfaces. We believe
that the point of view advocated in the present derivation provides valuable
physical insight to an otherwise rather unintuitive result. Secondly, we apply
the general formula of the multi-mode Hamiltonian for Brillouin optomechanics
to the important special case of a cylindrical nanowaveguide, and evaluate
analytically the parameters in the Hamiltonian characterizing the
phonon-photon coupling strength. We use our analytical expressions to
demonstrate the domination of radiation pressure effect over electrostriction for nanoscale waveguides,
and determine optimal parameter regimes exhibiting maximal coupling. The
formalism is applicable to systems of any dimensional scales, ranging from cavity
optomechanical systems involving localized photon and phonon modes up to bulk
materials in which photons and phonons are described by continuous fields as
in Brillouin optomechanics.

We treat a cylindrical nanoscale waveguide (tapered fiber), which exhibits a
(quasi)continuum of modes propagating in the longitudinal direction with
localized discrete modes due transverse confinement. We consider in much detail the case of a nanofiber made of silicon material that is embedded in free space. We analytically solve for the photon and phonon vector mode functions and dispersion relations and calculate the photon-phonon
coupling parameters originating from electrostriction and radiation pressure mechanisms. This case is motivated by the recent progress in Brillouin optomechanics \cite{Pant2011,Shin2013,VanLaer2015a,VanLaer2015b,Kittlaus2015}, but also by the work on  tapered optical nanofibers that have been used in manipulating, trapping, and detecting neutral cold and
ultracold atoms \cite{Vetsch2010,Goban2012}, in which optically active
mechanical modes of tapered nanofibers have been also investigated
\cite{Wuttke2013}. Moreover, we provide a means to compare the theoretically calculated photon-phonon
coupling parameter with the experimentally observed gain parameter. Moving to the real-space representation of the Hamiltonian, we solve the system of equations for the Stokes stimulated Brillouin scattering. The Stokes field is amplified with a gain parameter that is related to the photon-phonon coupling parameter in the derived Hamiltonian, similar to what has been discussed by Van Laer et
al. \cite{VanLaer2015c}.

The paper is organized as follows. Section \ref{sec:PerturbationTheory} contains the derivation of the
classical, perturbed Hamiltonians for the coupled light and mechanical
excitations through radiation pressure and electrostriction
mechanisms. The discussion is followed by the canonical quantization of the coupled classical fields, where the interacting multi-mode photon and phonon Hamiltonian is derived. The
photon-phonon coupling parameters are explicitly calculated for the case of cylindrical
nanowires in Section \ref{sec:Nanonwire}. A relation between the photon-phonon coupling parameter and the
gain parameter of Stokes stimulated Brillouin scattering is presented in Section \ref{sec:Gain}, and the coupled photon-phonon real-space Hamiltonian is
introduced. The appendices
include the full solutions of the electromagnetic fields and the mechanical
excitations in a cylindrical dielectric waveguide.

\section{Coupling among Electromagnetic Field and Mechanical Vibrations}\label{sec:PerturbationTheory}

We aim first to derive the classical Hamiltonian that represents the mutual influence
of the mechanical excitations and the electromagnetic field in a bounded
dielectric medium, and which are characterized by the electric field ${\bf E}({\bf x})$ and
the mechanical displacement field ${\bf Q}({\bf x})$, respectively. The electromagnetic
field in a dielectric, lossless, and nonmagnetic medium with a scalar permittivity $\epsilon({\bf x})$ is described by the Hamiltonian \cite{Glauber1991}
\begin{equation} \label{EMH}
H=\frac{1}{2}\int \mathrm{d}V\left\{\frac{1}{\epsilon({\bf x})}|{\bf D}({\bf x})|^2+\frac{1}{\mu_0}|{\bf
  B}({\bf x})|^2\right\},
\end{equation}
where the electric displacement field is defined by ${\bf D}({\bf
  x})=\epsilon({\bf x}){\bf E}({\bf   x})$. The Hamiltonian coupling of
photons and phonons follows from evaluating the correction $\delta H$ to $H$
due to a mechanical displacement ${\bf Q}({\bf x})$ causing a perturbation
$\delta\epsilon({\bf x})$ in the permittivity $\epsilon({\bf x})$. We will
consider corrections in first order of the mechanical displacement throughout
the paper, and limit the discussion to isotropic fluctuations in the dielectric constant. The explicit dependence of $\delta\epsilon({\bf x})$ on ${\bf Q}({\bf x})$ differs for radiation pressure and electrostrictive interaction, and will be detailed below. Nevertheless, both effects are covered  by the perturbation to the field Hamiltonian in Eq.~\eqref{EMH}
\begin{equation} \label{deltaEMH}
\delta H=\frac{1}{2}\int \mathrm{d}V\,
\delta\epsilon^{-1}({\bf x}) |{\bf D}({\bf x})|^2.
\end{equation}
Here $\delta\epsilon^{-1}({\bf x})$ in Eq.~(\ref{deltaEMH}) denotes the
perturbation of the inverse of the permittivity which is not to be confused
with the inverse of the perturbation $[\delta\epsilon({\bf x})]^{-1}$. The contributions to $\delta H$ due to the corrections in the amplitude of the electric displacement field ${\bf D}$ are negligibly small, on the order of the ratio of phonon to photon frequency, as shown in Appendix~\ref{sec:Maxwell}. A further correction of the same order is introduced through magnetic polarization effects \cite{Sipe2016}. Note that the
perturbation of the field Hamiltonian~(\ref{EMH}) needs to be done in the
representation where the energy density is expressed in terms of the electric
displacement field. If instead the electric field ${\bf E}$ is used, the
contributions due to corrections of the field amplitude will be significant, cf. Appendix~\ref{sec:Maxwell}, and $\delta H$ would take a much
more cumbersome form.

We consider a dielectric material with permittivity $\epsilon({\bf x})=\epsilon_1$ in a volume $V_1$, that is localized in a surrounding medium with permittivity $\epsilon({\bf x})=\epsilon_2$ occupying the complementary volume $V_2$. For a dielectric material in vacuum $\epsilon_2=\epsilon_0$.   The unperturbed permittivity can be written as
\begin{equation} \label{Dialectric}
\epsilon({\bf
  x})=\epsilon_2+(\epsilon_1-\epsilon_2)\Theta({\bf x}),
\end{equation}
where $\Theta({\bf x})$ is a step function defined by
\begin{align*}
  \Theta({\bf   x})&=\begin{cases} 1 & \mathrm{for}  \quad{\bf x} \in V_1\\  0 & \mathrm{for}  \quad{\bf x}\in V_2\end{cases}  .
\end{align*}
A mechanical displacement ${\bf Q}({\bf x})$ of the dielectric medium in $V_1$ will affect the material's permittivity in two ways:  In the electrostrictive mechanism, the fluctuations change the magnitude of the permittivity of the material in $V_1$ at a fixed boundary. Radiation pressure in turn corresponds to fluctuations in the material boundary at a fixed magnitude of $\epsilon_1$. Thus, the perturbed permittivity can be written as
\[
\epsilon({\bf   x})=\epsilon_2+\left[\epsilon_1({\bf Q})-\epsilon_2\right]\Theta({\bf x+{\bf Q}}),
\]
where $\epsilon_1({\bf Q})|_{{\bf Q}=0}=\epsilon_1$. The first order corrections is $\delta\epsilon({\bf   x})=\delta\epsilon_\mathrm{rp}({\bf   x})+\delta\epsilon_\mathrm{el}({\bf   x})$ and the contributions from radiation pressure and electrostriction are, respectively,
\begin{align}
\delta\epsilon_\mathrm{rp}({\bf   x})&=(\epsilon_1-\epsilon_2){\bf Q}({\bf
  x})\cdot\nabla\Theta({\bf x})\label{eq:deltaRP},\\
\delta\epsilon_\mathrm{el}({\bf   x})&=\delta\epsilon_1({\bf Q})\Theta({\bf x}).\label{eq:deltaepsEL}
\end{align}
The contributions of these two perturbations to the interaction Hamiltonian in Eq.~\eqref{deltaEMH} will be treated in the following two subsections.

\subsection{Radiation Pressure}

We note first that $\nabla\Theta({\bf x})$ in Eq.~\eqref{eq:deltaRP} denotes a delta distribution whose effect is to turn volume integrals into surface integrals over the boundary $\partial V$ between $V_1$ and $V_2$, that is
\begin{align}\label{eq:theta}
\int_{V_1+V_2} \mathrm{d}V\, {\bf f}({\bf x})\cdot \nabla\Theta({\bf x})=\int_{\partial V}\mathrm{d}{\bf A}\cdot {\bf f}({\bf x}),
\end{align}
where $\mathrm{d}{\bf A}$ is the infinitesimal surface element.  Thus, when using Eq.~\eqref{eq:deltaRP} in \eqref{deltaEMH} it will be vital to take care that the remaining integrand does not contain discontinuities on the boundary surface rendering the integral undefined.

This concerns in particular the discontinuity in the field component of ${\bf D({\bf x})}$ parallel to the boundary surface. In order to overcome this difficulty we consider an (arbitrarily thin) shell volume ${V^\prime}$  enclosing the boundary surface $\partial V$, and rewrite the contribution to the interaction Hamiltonian \eqref{deltaEMH} within $V^\prime$  in terms of the fields ${\bf E}_{\parallel}({\bf x})={\bf D}_{\parallel}({\bf x})/\epsilon({\bf x})$ and ${\bf D}_{\perp}({\bf x})$ parallel and orthogonal to the surface that are both continuous,
\begin{equation} \label{EMH_shell}
\delta H=\frac{1}{2}\int_{V^\prime} \mathrm{d}V\left\{-\delta\epsilon({\bf x})|{\bf E}_{\parallel}({\bf x})|^2+\delta\epsilon^{-1}({\bf x})|{\bf D}_{\perp}({\bf x})|^2\right\}.
\end{equation}
Here we used $\epsilon^{2}({\bf x})\delta\epsilon^{-1}({\bf x})=-\delta \epsilon({\bf x})$. In restricting the integration to the shell volume $V^\prime$ we anticipate, in view of Eq.~\eqref{eq:theta}, that it is the energy within this subvolume which will be relevant for the perturbation due to radiation pressure. The remaining hurdle is to arrive at a well behaved perturbation of the inverse permittivity in Eq.~\eqref{EMH_shell}. This can be achieved by expressing the inverse as $\epsilon^{-1}({\bf  x})=\epsilon_2^{-1}+(\epsilon_1^{-1}-\epsilon_2^{-1})\Theta({\bf x})$ which yields
\begin{equation}\label{eq:deltaepsRP_inv}
\delta\epsilon_\mathrm{rp}^{-1}({\bf x})=\left(\epsilon^{-1}_1-\epsilon^{-1}_2\right) {\bf Q}({\bf
  x})\cdot\nabla\Theta({\bf x}),
\end{equation}
 for the contribution due to radiation pressure.

Finally, we can use Eqs.~\eqref{eq:deltaRP}, \eqref{eq:theta}, and \eqref{eq:deltaepsRP_inv} in expression \eqref{EMH_shell} to arrive at the Hamiltonian describing radiation pressure interaction
\begin{eqnarray}\label{HamRP}
\delta H_\mathrm{rp}&=&-\frac{1}{2}\int_{\partial V} \mathrm{d}{\bf A}\cdot{\bf
  Q}({\bf x}) \nonumber \\
&\times&\left\{\Delta\epsilon|{\bf
  E}_{\parallel}({\bf
  x})|^2-\Delta(\epsilon^{-1})|{\bf D}_{\perp}({\bf x})|^2\right\}.
\end{eqnarray}
We used the symbols $\Delta\epsilon=\epsilon_1-\epsilon_2$ and $\Delta(\epsilon^{-1})=\epsilon^{-1}_1-\epsilon^{-1}_2$ following the notation introduced by Johnson et al. in \cite{Johnson2002}. The integral in Eq.~\eqref{HamRP} is over the surface $\partial V$ of the dielectric medium, and all fields are evaluated on that surface. Thanks to the continuity of ${\bf E}_{\parallel}$ and ${\bf D}_{\perp}$ there is no ambiguity in evaluating the field on either side of the surface.

The result presented here agrees with the one derived in \cite{Johnson2002} for the perturbed eigenfrequency of photonic field modes. Johnson et al. used a smoothing procedure in order to deal with the difficulty of discontinuities of field amplitudes at the surface. The alternative approach presented here avoids such technicalities and at the same time provides directly the interaction Hamiltonian covering both frequency shifts of photon modes and Brillouin scattering among different field modes. It is this last aspect which is of main interest in the description of the optomechanics of extended nanophotonic waveguides. Furthermore, the interaction Hamiltonian \eqref{HamRP}  is directly amenable to quantization, as will be done in Sec.~\ref{sec:quantization}, and provides firm grounds for the description of  radiation pressure effects in Brillouin quantum optomechanics.

\subsection{Electrostriction}

Electrostriction appears due to the tendency of a dielectric material to be
compressed in the presence of light, and as a consequence to excite mechanical vibrations in the medium \cite{Fabelinskii1968,Boyd2008}. The appearance of mechanical vibrations modulate the optical properties and result in small changes in the dielectric constant that induce scattering of the light. A Hamiltonian description of electrostriction has been derived previously by Agrawal et al.~\cite{Agarwal2013a}. For completeness we  present a derivation within the present approach, starting from expression~\eqref{eq:deltaepsEL} for the perturbation of the bulk value of the permittivity due to a mechanical displacement.

By means of the electrostriction constant $\gamma_\mathrm{el}$ the change in the
permittivity can be related to fluctuations of the mass density $\rho$
\begin{equation}
\delta\epsilon_1({\bf Q})=\epsilon_0\frac{\gamma_\mathrm{el}}{\rho}\delta\rho({\bf Q}),
\end{equation}
which in turn is determined through the mechanical displacement
\begin{equation}
\delta\rho({\bf Q})\simeq -\rho\nabla\cdot{\bf Q}({\bf x}).
\end{equation}
Overall, we arrive from Eq.~(\ref{eq:deltaepsEL}) at
\begin{equation} \label{eq:deltaepsEL}
\delta\epsilon_\mathrm{el}({\bf x})=-\epsilon_0\gamma_\mathrm{el}(\nabla\cdot{\bf Q}({\bf x}))\Theta({\bf x}).
\end{equation}
Electrostriction mechanism in nanoscale structures can give rise to
anisotropic phenomena. Anisotropic contributions induced by longitudinal
phonons in nanoscale waveguides are immaterial \cite{Qiu2013}. As we
concentrate mainly in processes involving longitudinal phonons we treat here
only the isotropic case of scalar fluctuations \cite{Fabelinskii1968,Boyd2008}. Anisotropic phenomena
induced by tensor fluctuations of the dielectric function are beyond the scope
of the present paper.

The relation (\ref{eq:deltaepsEL}) can be used directly in Eq.~(\ref{deltaEMH}) when resorting to the
representation of $\delta H$ in terms of the electric field (using again
$\epsilon^2({\bf x})\delta(\epsilon^{-1}({\bf x}))=-\delta\epsilon({\bf x})$)
\begin{align}
\delta H_\mathrm{el}=-\frac{1}{2}\int_{V_1+V_2}
\mathrm{d}V\delta\epsilon_\mathrm{el}({\bf x})|{\bf E}({\bf x}))|^2.
\end{align}
No issues regarding discontinuities in the integrand arise here since the domain of integration is effectively limited to the volume $V_1$ occupied by
the dielectric due to the step function in Eq.~(\ref{eq:deltaepsEL}). Here
$\delta H_\mathrm{el}$ decreases by increasing $\delta\epsilon$ as
expected. Finally, this yields the Hamiltonian for the electrostrictive
interaction of photons and phonons,
\begin{align}\label{HamEL}
\delta H_\mathrm{el}&=\gamma_\mathrm{el}\frac{\epsilon_0}{2}\int_{V_1}\mathrm{d}V(\nabla\cdot{\bf Q}({\bf x}))|{\bf E}({\bf x}))|^2.
\end{align}
It is evident from Eqs.~(\ref{HamRP}) and (\ref{HamEL}) that radiation
pressure and electrostriction are surface and volume effects,
respectively. For sufficiently small dimensions radiation pressure will
therfore dominate, and may ultimately provide largely enhanced coupling
strengths per single photon and phonon. We will show this explicitly for the
example of a cylindrical waveguide in Sec.~\ref{sec:Nanonwire}. Before that,
we will quantized the interaction Hamiltonians in Eqs.~(\ref{HamRP}) and
(\ref{HamEL}), and extract the quantum mechanical coupling strengths at the
single photon/phonon level.

\subsection{Quantization}\label{sec:quantization}

We aim to derive the photon-phonon interaction Hamiltonian in dielectric media by canonically quantizing the electromagnetic and mechanical fields that were presented in the previous Sections. The Hamiltonian of the free photon field reads \cite{Glauber1991,Sipe2004}
\begin{equation}
\hat{H}_\mathrm{phot}=\sum_{\alpha}\hbar\omega~\hat{a}_{\alpha}^{\dagger}\hat{a}_{\alpha},
\end{equation}
where the summation is over all the photon modes and the summation index $\alpha$ comprises any index labeling photon modes for a given geometry. In the following we will assume a discrete set of indices, such that $\hat{a}_{\alpha}^{\dagger}$ and $\hat{a}_{\alpha}$ are dimensionless bosonic creation and annihilation
operators, and fulfill
$[\hat{a}_\alpha,\hat{a}^\dagger_\beta]=\delta_{\alpha\beta}$. Continuous index sets as relevant to problems with \textit{e.g.} translational symmetry can be attained via appropriate limiting procedures, as detailed for example in Ref.~\cite{Loudon2000}.  The electric field operator reads
$\hat{\bf E}({\bf x})=\hat{\bf E}^{(+)}({\bf x})+\hat{\bf E}^{(-)}({\bf
  r})$, where
\begin{equation}
\hat{\bf
  E}^{(+)}({\bf
  r},t)=i\sum_{\alpha}{\cal E}_{\alpha}\hat{a}_{\alpha}{\bf
  U}_{\alpha}({\bf x}),
\end{equation}
and $\hat{\bf E}^{(-)}=(\hat{\bf E}^{(+)})^{\dagger}$. The (dimensionless) vector function ${\bf U}_{\alpha}({\bf x})$ of mode $\alpha$ and the eigenfrequency $\omega_\alpha$ are obtained by solving Maxwell's equations with appropriate boundary conditions, cf. Appendix~\ref{sec:Maxwell}. The electric field
amplitude of a single photon in mode $\alpha$ is
\begin{equation}
{\cal E}_{\alpha}=\sqrt{\frac{\hbar\omega_\alpha}{2\epsilon_0V_{\alpha}^\mathrm{phot}}}.
\end{equation}
 The effective volume $V_{\alpha}^\mathrm{phot}$ is calculated from the normalization condition
\begin{equation}
\int \mathrm{d}V|{\bf
  U}_{\alpha}({\bf x})|^2=V_{\alpha}^\mathrm{phot}.
\end{equation}

On the other hand, the phonon Hamiltonian reads
\begin{equation}
\hat{H}_\mathrm{phon}=\sum_{\nu}\hbar\Omega_{\nu}~\hat{b}_{\nu}^{\dagger}\hat{b}_{\nu},
\end{equation}
where $\hat{b}_{\nu}^{\dagger}$ and $\hat{b}_{\nu}$ are the bosonic creation and annihilation
operators of a phonon in mode $\nu$ with angular frequency $\Omega_{\nu}$. The index $\nu$ again stands for any discrete index relevant for a given geometry. The
 operator corresponding to the mechanical displacement field is $\hat{\bf Q}({\bf x})=\hat{\bf Q}^{(+)}({\bf x})+\hat{\bf Q}^{(-)}({\bf x})$, where
\begin{equation}
\hat{\bf Q}^{(+)}({\bf x})=\sum_{\nu}Z_{\nu}\hat{b}_{\nu}{\bf W}_{\nu}({\bf x}),
\end{equation}
and $\hat{\bf Q}^{(-)}=(\hat{\bf Q}^{(+)})^{\dagger}$. Here ${\bf W}_{\nu}({\bf x})$ is the (dimensionless) vector mode function of phonons with eigenfrequency $\Omega_{\nu}$ which is obtained by solving the equations of motion of mechanical excitation as shown in Sec.~\ref{sec:Nanonwire}. The
zero-point-fluctuation of mode $\nu$ is
\begin{equation}
Z_{\nu}=\sqrt{\frac{\hbar}{2M^\mathrm{phon}_{\nu}\Omega_{\nu}}}.
\end{equation}
The effective mass associated to phonon mode $\nu$ is given by $M_{\nu}^\mathrm{phon}=\rho V_{\nu}^\mathrm{phon}$, where $\rho$ is the
medium's mass density and $V_{\nu}^\mathrm{phon}$ is the phonon
effective volume, which can be calculated from the normalization condition
\begin{equation}
\int \mathrm{d}V|{\bf W}_{\nu}({\bf x})|^2=V_{\nu}^\mathrm{phon}.
\end{equation}

The quantized interaction Hamiltonian is obtained from the classical one
 by replacing the displacement and the electromagnetic fields by operators, and
using normal ordering. We will use the notation $({\bf :}\hat{X}{\bf :})$ to denote the normally ordered form of the operator $\hat{X}$. The radiation pressure Hamiltonian reads
\begin{eqnarray}
\hat{H}_\mathrm{rp}&=&-\frac{1}{2}\int_{\partial V}{\bf :}\mathrm{d}{\bf
  A}\cdot\hat{\bf Q}({\bf x})\nonumber \\
&\times&\left\{\Delta\epsilon|\hat{\bf
  E}_{\parallel}({\bf
  x})|^2-\Delta(\epsilon^{-1})|\hat{\bf D}_{\perp}({\bf x})|^2\right\}{\bf :}
\end{eqnarray}
and the electrostriction Hamiltonian reads
\begin{equation}
\hat{H}_\mathrm{el}=\gamma_\mathrm{el}\frac{\epsilon_0}{2}\int_{V_1} dV{\bf :}\left(\nabla\cdot \hat{\bf
  Q}({\bf x})\right)|\hat{\bf E}({\bf x})|^2{\bf :}
\end{equation}
In terms of creation and annihilation operators, using the above definitions, the
interaction Hamiltonian is
\begin{eqnarray}\label{eq:HintQ}
\hat{H}_{I}&=&\hat{H}_\mathrm{rp}+\hat{H}_\mathrm{el}\nonumber \\
&=&\hbar\sum_{\alpha\alpha'\nu}\left\{f^{\ast}_{\alpha\alpha'\nu}~\hat{b}_{\nu}^{\dagger}\hat{a}_{\alpha'}^{\dagger}\hat{a}_{\alpha}+f_{\alpha\alpha'\nu}~\hat{a}_{\alpha}^{\dagger}\hat{a}_{\alpha'}\hat{b}_{\nu}\right\}.
\end{eqnarray}
Thus, the optomechanical photon-phonon coupling strength (of dimension Hz) is $f_{\alpha\alpha'\nu}=f_{\alpha\alpha'\nu}^\mathrm{rp}+f_{\alpha\alpha'\nu}^\mathrm{el}$, where the radiation pressure coupling is given
by
\begin{align}
&f_{\alpha\alpha'\nu}^\mathrm{rp}=-\frac{1}{2}Z_{\nu}{\cal E}_{\alpha}{\cal
  E}_{\alpha'}\int_{\partial V}\mathrm{d}{\bf A}\cdot{\bf W}_{\nu}({\bf
    x})~\times \nonumber \\
&\left\{\Delta\epsilon\
{\bf U}_{\alpha}^{\parallel\ast}({\bf x})\cdot{\bf U}_{\alpha'}^{\parallel}({\bf x})-\epsilon_1^2\Delta(\epsilon^{-1}){\bf U}_{\alpha}^{\perp<\ast}({\bf x})\cdot{\bf U}_{\alpha'}^{\perp<}({\bf x})\right\}.
\end{align}
All mode functions are evaluated on the boundary surface, and most
important here is that the perpendicular components, ${\bf U}_{\alpha}^{\perp<}$, are
evaluated on the internal side of the surface, as indicated by the superscript
symbol $(<)$. For the electrostrictive coupling we get
\begin{equation}
f_{\alpha\alpha'\nu}^\mathrm{el}=\gamma_{el}\frac{\epsilon_0}{2}Z_{\nu}{\cal E}_{\alpha}{\cal
  E}_{\alpha'}\int_{V_1} dV\left(\nabla\cdot {\bf
  W}_{\nu}({\bf x})\right){\bf U}_{\alpha}^{\ast}({\bf x})\cdot{\bf U}_{\alpha'}({\bf x}).
\end{equation}
The results are applicable to any dielectric material ranging
from fully confined media, as for a resonator, up to partly confined media, as
for a waveguide. The interactions are consistent with the results
in \cite{Sipe2016}, but here we give explicitly the appropriate normalized amplitudes.

\section{Coupling among Photons and Phonons in Nanophotonic Waveguides}\label{sec:Nanonwire}

\begin{figure}
\includegraphics[width=.6\columnwidth]{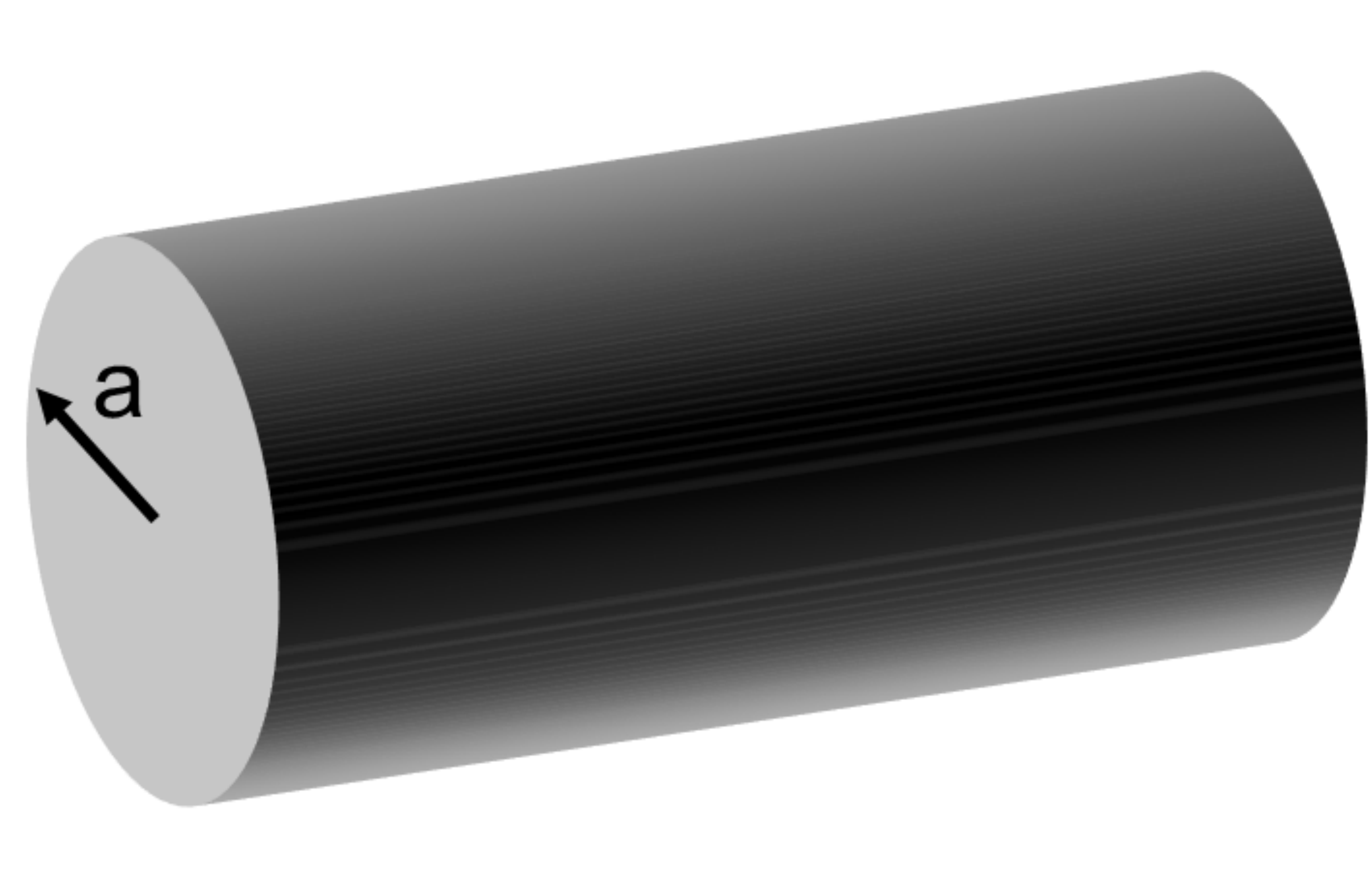}
\caption{An optical nanofiber of radius $a$ and refractive index $n$, which is
  localized in free space.}
\label{Fiber}
\end{figure}

The formalism of the previous section has been applied extremely successfully
to quantum optomechanical systems comprising single electromagnetic and
mechanical modes. Our main concern here is the application to extended media,
where the multimode character of the interaction Hamiltonian is essential but
at the same time field confinement due to small spatial extensions yields
strong coupling due to enhanced radiation pressure. In the following we
present analytical results for the coupling strengths
$f_{\alpha\alpha'\nu}^\mathrm{rp}$ and  $f_{\alpha\alpha'\nu}^\mathrm{el}$ for
the most elementary such geometry, namely a cylindrical nanophotonic
waveguide. We consider a dielectric nanowire that is localized in free space
and extended along the $z$ direction over a length $L$ with a nanoscale radius
$a$, as seen in figure (\ref{Fiber}). The dielectric constants are
$\epsilon_1=\epsilon_0 n^2$ and $\epsilon_2=\epsilon_0$, where $n$ is the medium refractive index. Recently, such tapered optical nanofibers have been intensively studied \cite{Vetsch2010,Goban2012,Wuttke2013}.

The strong confinement in the transverse direction results in discrete modes for both the electromagnetic and mechanical fields. In the following we will consider only a single transverse mode for the photons and phonons. The electromagnetic and mechanical fields can propagate along the waveguide axis with wavenumber
$k$ which takes on the values $k=\frac{2\pi}{L}m$ with $m=0,\pm1,\pm2,\ldots$, where $L$ is the waveguide length and we use periodic boundary conditions. Phonon wavenumbers will be denoted by $q$, and have a natural cut-off that is given by the inverse of the crystal lattice constant.

\subsection{Photons in Nanoscale Waveguides}\label{sec:phot}

The photon Hamiltonian for a single tranverse mode reads
\begin{equation}\label{eq:Hphot}
\hat{H}_\mathrm{phot}=\sum_{k}\hbar\omega_{k}~\hat{a}_k^{\dagger}\hat{a}_k,
\end{equation}
where $\hat{a}_k^{\dagger}$ and $\hat{a}_k$ are the creation and annihilation
operators of a photon of wavenumber $k$ and angular frequency $\omega_k$. The electric field operator reads, in cylindrical coordinates,
\begin{align}\label{EEE}
\hat{\bf
  E}(r,\theta,z)&=i\sum_{k}\sqrt{\frac{\hbar\omega_k}{2\epsilon_0V_{k}^\mathrm{phot}}} \nonumber \\
&\times\left\{\hat{a}_k{\bf
  u}_k(r,\theta)e^{ikz}-\hat{a}_k^{\dagger}{\bf
  u}_k^{\ast}(r,\theta)e^{-ikz)}\right\},
\end{align}
where ${\bf u}_k(r,\theta)$ is the transverse vector mode function, and the effective mode volume $V_{k}^\mathrm{phot}$ is defined by
$\int \mathrm{d}V|{\bf
  u}_k(r,\theta)|^2=V_k^\mathrm{phot}
$.

\begin{figure}
\includegraphics[width=0.8\columnwidth]{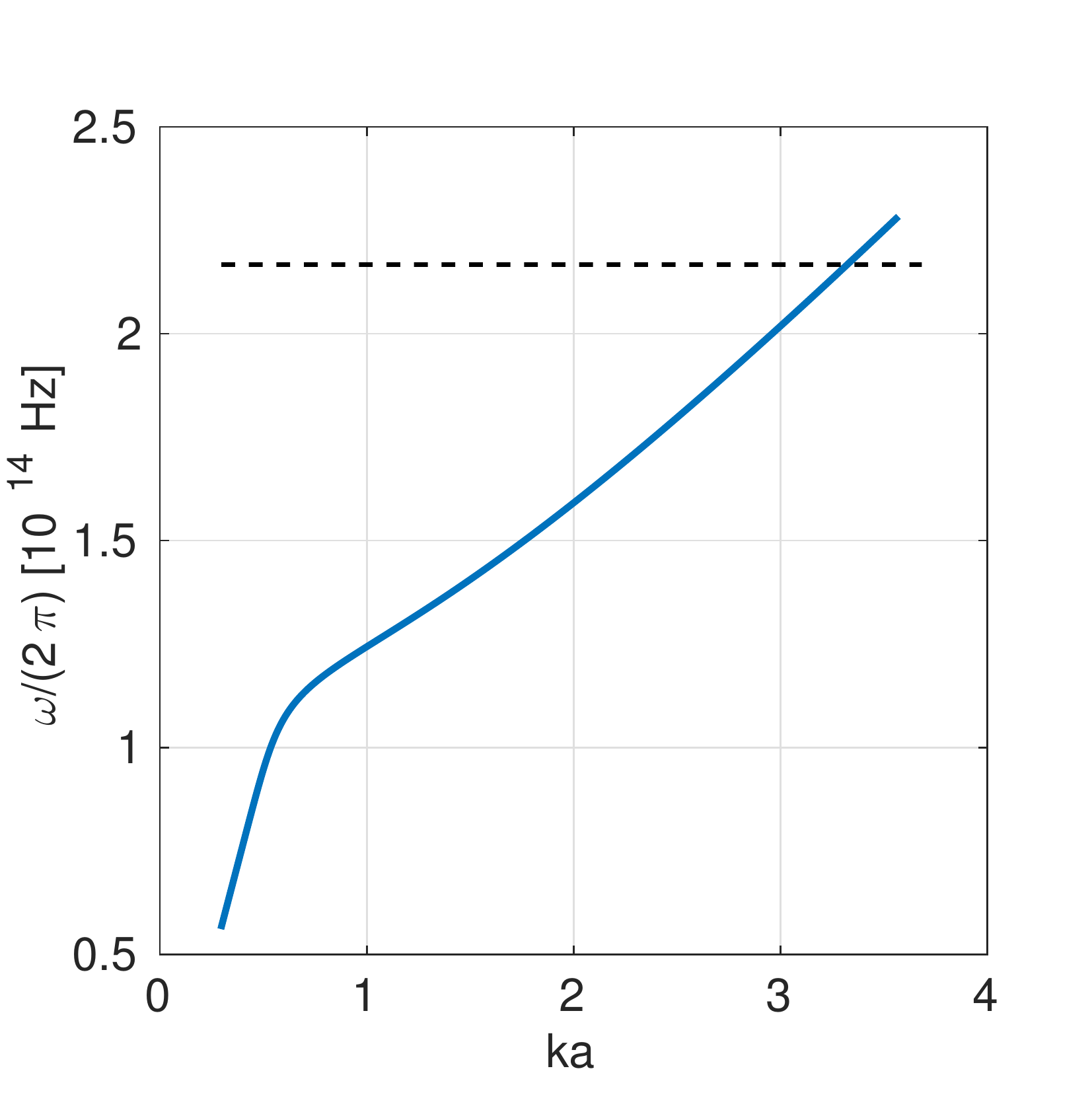}
\caption{The lowest HE$_{11}$ fiber mode is presented as
  $\omega/(2\pi)$ vs. $ka$. Here $n=3.5$ for silicon material. Beyond the dashed line at $\omega/(2\pi)\approx
  2.186\times 10^{14}$ Hz other photon branches start to appear.}
\label{PhotDis}
\end{figure}

The lowest propagating mode in an optical nanofiber is the HE$_{11}$ mode. We concentrate in photons with a fixed polarization, and we consider rotating polarization with left or right hand circular rotations. Detailed
calculations of the cylindrical waveguide photon dispersions and mode functions
are given in Appendix~\ref{App:FibreField}. The
photon dispersion, which is the relation between the angular frequency,
$\omega$, and the wavenumber along the fiber
axis, $k$, can be extracted from the expression \cite{Kien2004}
 \begin{align}
\frac{J_{0}(pa)}{paJ_1(pa)}&=\left(\frac{1+n^2}{2n^2}\right)\frac{K_{0}(qa)+K_{2}(qa)}{2qaK_1(qa)}+\frac{1}{p^2a^2} \nonumber \\
\quad&-\left\{\left(\frac{n^2-1}{2n^2}\right)^2\left(\frac{K_{0}(qa)+K_{2}(qa)}{2qaK_1(qa)}\right)^2\right. \nonumber \\
&+\left.\left(\frac{k}{nk_0}\right)^2\left(\frac{1}{q^2a^2}+\frac{1}{p^2a^2}\right)^2\right\}^{1/2},
\end{align}
where $p=\sqrt{k_0^2n^2-k^2}$, and $q=\sqrt{k^2-k_0^2}$, with
$k_0=\omega/c$. Propagating modes can appear only in the range
$1\leq\frac{k}{k_0}\leq n$. In the literature this result is commonly
represented in terms of the
fundamental parameter $V=k_0a\sqrt{n^2-1}$. For silicon we have $n\approx 3.5$
and up to about $V\approx 3.84$ only the HE$_{11}$ photons propagate in the
fiber, and beyond $V\approx 3.84$ TM and TE modes can be excited. The HE$_{11}$ dispersion relation is plotted in Figure~\ref{PhotDis} for $\omega/(2\pi)$ as a function of $ka$. For small wavenumbers the
photons are unconfined in the nanofiber and propagate with the
group velocity $c/n$, but beyond $ka\approx 0.7$ they get confined and propagate
with almost linear dispersion of group velocity $v_g\approx c/5$.

\begin{figure}
\includegraphics[width=0.8\columnwidth]{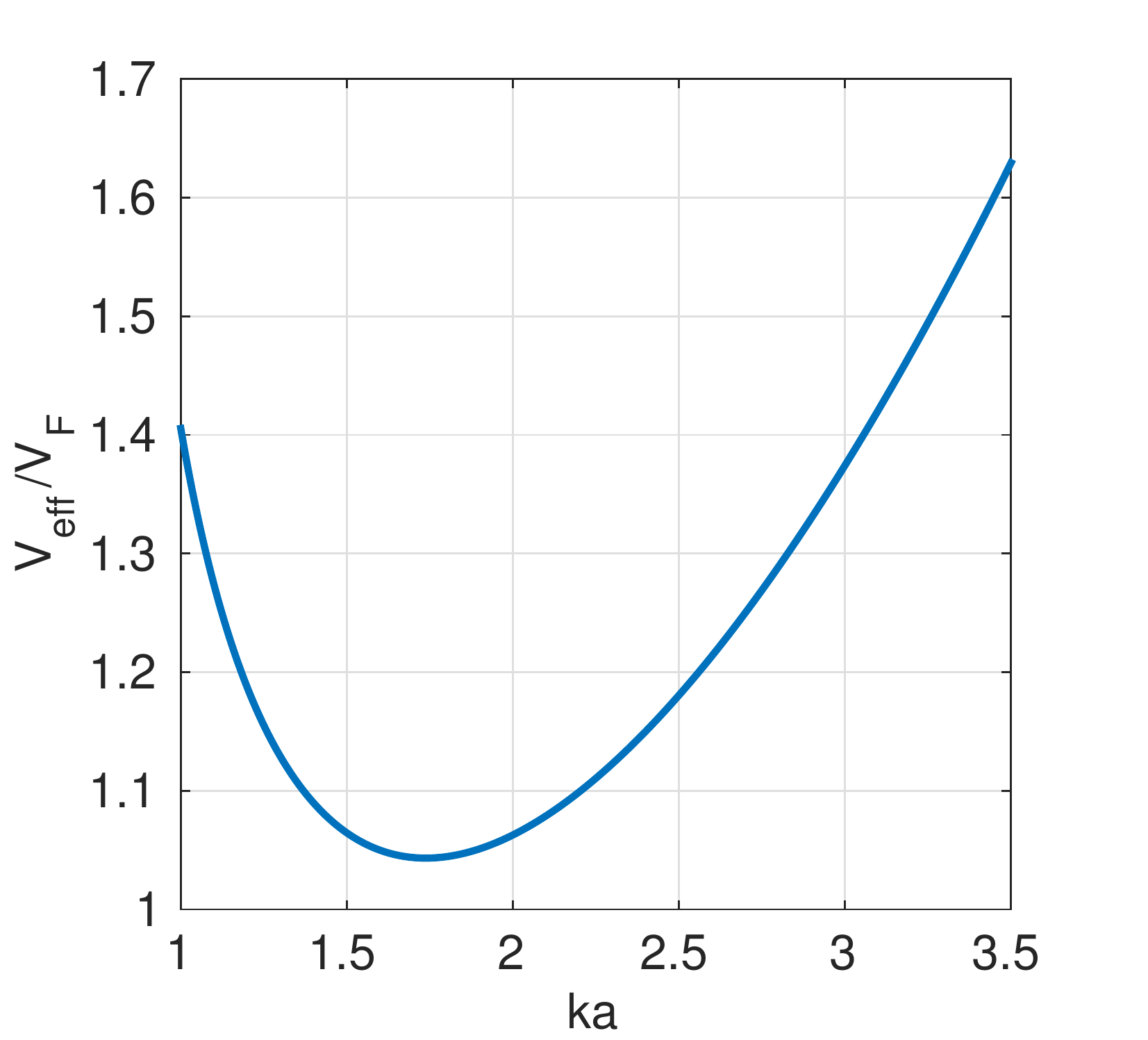}
\caption{The relative photon mode volume $V_{k}^\mathrm{phot}/V_F$ vs. $ka$ for HE$_{11}$ fiber photons.}
\label{VolPhot}
\end{figure}

The vector mode functions inside the fiber, that is for $(r<a)$, are given by
\begin{align}\label{eq:u_inner}
u^{r<}_k&=-iB\frac{k}{2p}\left[(1-s)J_0(pr)-(1+s)J_2(pr)\right]e^{\pm i\theta}, \nonumber \\
u^{\theta<}_k&=\pm B
\frac{k}{2p}\left[(1-s)J_0(pr)+(1+s)J_2(pr)\right]e^{\pm i\theta},\nonumber \\
u^{z<}_k&=BJ_1(pr)e^{\pm i\theta},
\end{align}
and outside the fiber, that is $(r>a)$, are given by
\begin{align}\label{eq:u_outer}
u^{r>}_k&=-iB\frac{k}{2q}\frac{J_1(pa)}{K_1(qa)} \nonumber \\
&\left[(1-s)K_0(qr)+(1+s)K_2(qr)\right]e^{\pm
  i\theta}, \nonumber \\
u^{\theta>}_k&=\pm B
\frac{k}{2q}\frac{J_1(pa)}{K_1(qa)} \nonumber \\
&\left[(1-s)K_0(pr)-(1+s)K_2(qr)\right]e^{\pm
  i\theta}, \nonumber \\
u^{z>}_k&=B\frac{J_1(pa)}{K_1(qa)}K_1(qr)e^{\pm i\theta},
\end{align}
where
\begin{equation}
s=\left[\frac{1}{p^2a^2}+\frac{1}{q^2a^2}\right]\left\{\frac{J_1^{\prime}(pa)}{paJ_1(pa)}+\frac{K_1^{\prime}(qa)}{qaK_1(qa)}\right\}^{-1},
\end{equation}
and $(\pm\theta)$ stand for left and right hand circular
polarizations. The parameter $B$ is fixed from the normalization relation stated above. In Figure~\ref{VolPhot} we plot the effective volume $V_{k}^\mathrm{phot}$ of photon mode $k$ relative to the total fiber volume $V_F=\pi a^2L$ as a
function of $ka$. Here a minimum appears around $ka\approx 1.74$ in
which the photon mode is highly concentrated inside the fiber, and a small
part penetrates outside. For example, for $a=250$~nm we get a minimum at $\lambda\approx 900$~nm.

\subsection{Phonons in Nanoscale Waveguides}\label{sec:phon}

The phonon Hamiltonian for a single mode is given by
\begin{equation}\label{eq:phonH}
\hat{H}_\mathrm{phon}=\sum_{q}\hbar\Omega_{q}~\hat{b}_{q}^{\dagger}\hat{b}_{q},
\end{equation}
where $\hat{b}_{q}^{\dagger}$ and $\hat{b}_{q}$ are the creation and annihilation
operators of a phonon of wavenumber $q$ and angular frequency $\Omega_{q}$. The displacement operator is defined by
\begin{equation}\label{QQQ}
\hat{\bf Q}(r,\theta,z)=\sum_{q}Z_q\left\{\hat{b}_{q}{\bf w}_{q}(r,\theta)e^{iqz}+\hat{b}_{q}^{\dagger}{\bf w}_{q}^{\ast}(r,\theta)e^{-iqz)}\right\},
\end{equation}
where ${\bf w}_{q}(r,\theta)$ is the transverse vector mode function. The
zero-point-fluctuation is $Z_q=(\hbar/2M_q\Omega_{q})^{1/2}$ with effective mass $M_q=\rho V_q^\mathrm{phon}$ and effective phonon mode
volume $\int \mathrm{d}V|{\bf w}_q(r,\theta)|^2=V_q^\mathrm{phon}$.

\begin{figure}
\includegraphics[width=0.8\columnwidth]{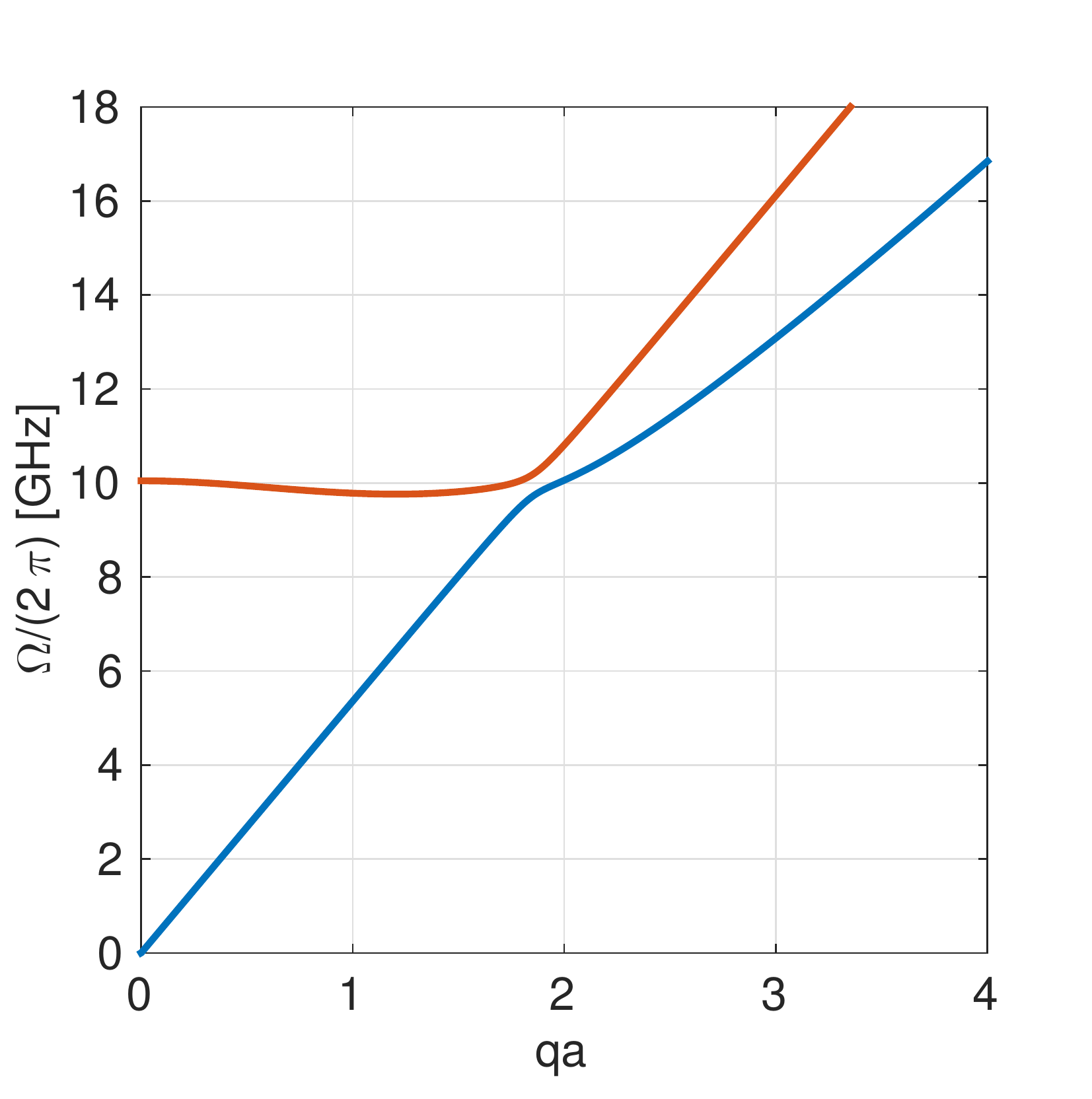}
\caption{The lowest two branches of the longitudinal phonon modes, for $\Omega/(2\pi)$
  vs. $qa$ for a silicon waveguide ($n=3.5$): Lower, acoustic branch (blue) and upper vibrational branch (red).}
\label{PhonDis}
\end{figure}

In optical nanofibers torsional, longitudinal and flexural phonons can be
excited. Here we consider only longitudinal modes, as the torsional modes
decouple to the light through radiation pressure and the flexural modes are of
higher energy. Detailed calculations of the cylindrical waveguide phonon dispersions and mode functions
are given in Appendix~\ref{App:FibrePhonons}. The longitudinal phonon dispersion can be extracted from the
expression \cite{Achenbach1975}
\begin{equation}
(q^2-\eta_t^2)^2\frac{\eta_laJ_0(\eta_la)}{J_1(\eta_la)}+4q^2\eta_l^2\frac{\eta_taJ_0(\eta_ta)}{J_1(\eta_ta)}=2\eta_l^2(q^2+\eta_t^2),
\end{equation}
where we have
$\eta_l^2=\frac{\Omega^2}{v_l^2}-q^2$, and
$\eta_t^2=\frac{\Omega^2}{v_t^2}-q^2$. The lowest two longitudinal branches are plotted in Figure~\ref{PhonDis} for $\Omega/(2\pi)$ as a function of $qa$. We treat silicon material
with $v_l=8433$~m/s and $v_t=5843$~m/s. For small wavenumbers $qa\ll 1$, the
lowest acoustic modes have a linear dispersion, and the lowest vibrational modes are
almost dispersion-less up to $ka\approx 2$. Both branches become linear beyond
the anti-crossing point.

The vector mode functions are given by
\begin{align} \label{PhonAmpl}
w^r_q&=-A\eta_lJ_1(\eta_l
r)+iCqJ_1(\eta_tr), \nonumber \\
w^{\theta}_q&=0, \nonumber \\
w^z_q&=iAqJ_0(\eta_l r)-C\eta_tJ_0(\eta_tr).
\end{align}
The parameters $A$ and $C$ are fixed using the boundary condition relation
\begin{equation}
C=\frac{2iq\eta_l}{\left[\eta_t^2-q^2\right]}\frac{J_1(\eta_la)}{J_1(\eta_ta)}A,
\end{equation}
and the normalization relation stated above. Note that $A$
also plays the role of a scaling factor that takes care of the appropriate
units.

\begin{figure}
\includegraphics[width=0.6\columnwidth]{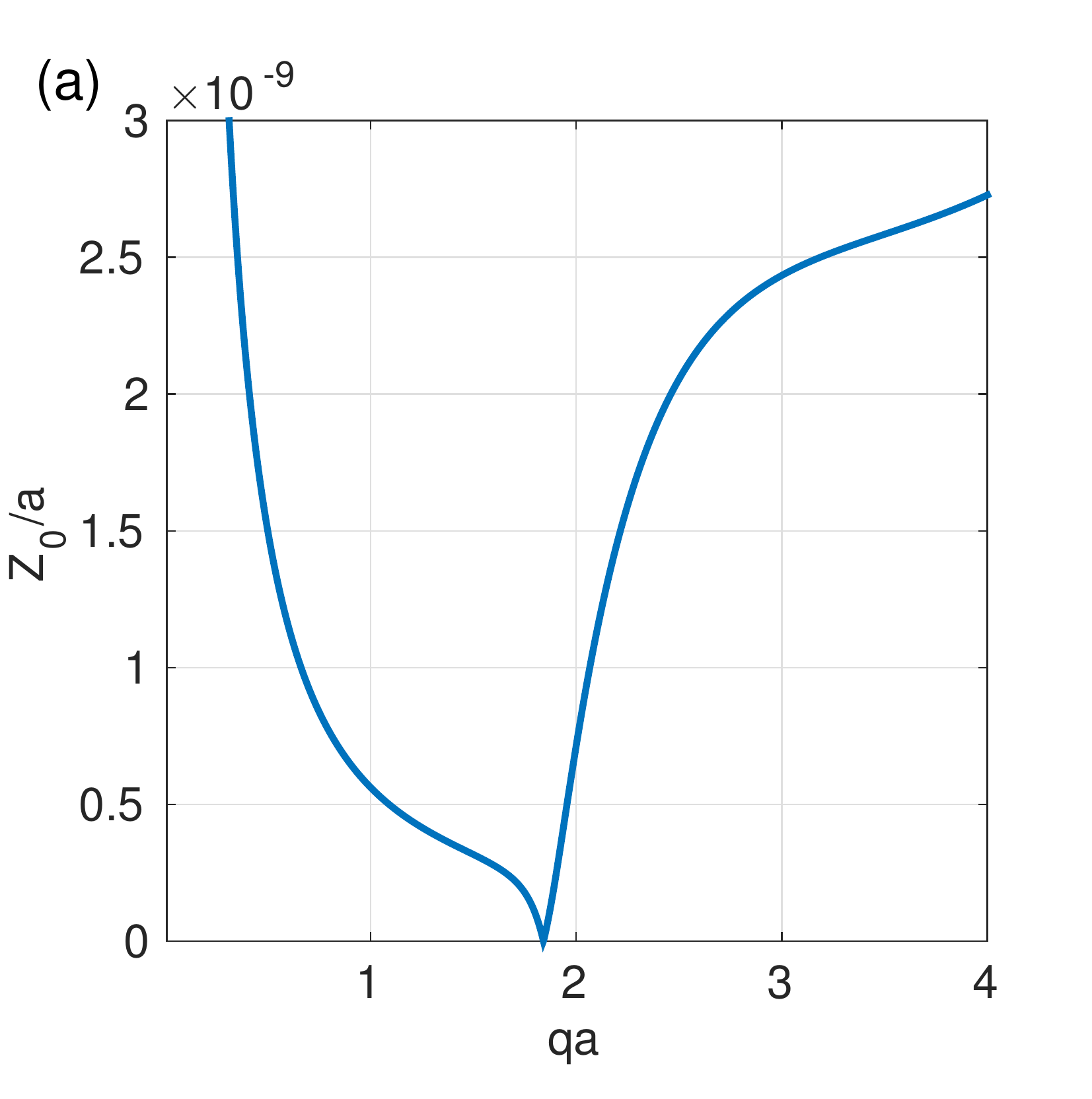}
\includegraphics[width=0.6\columnwidth]{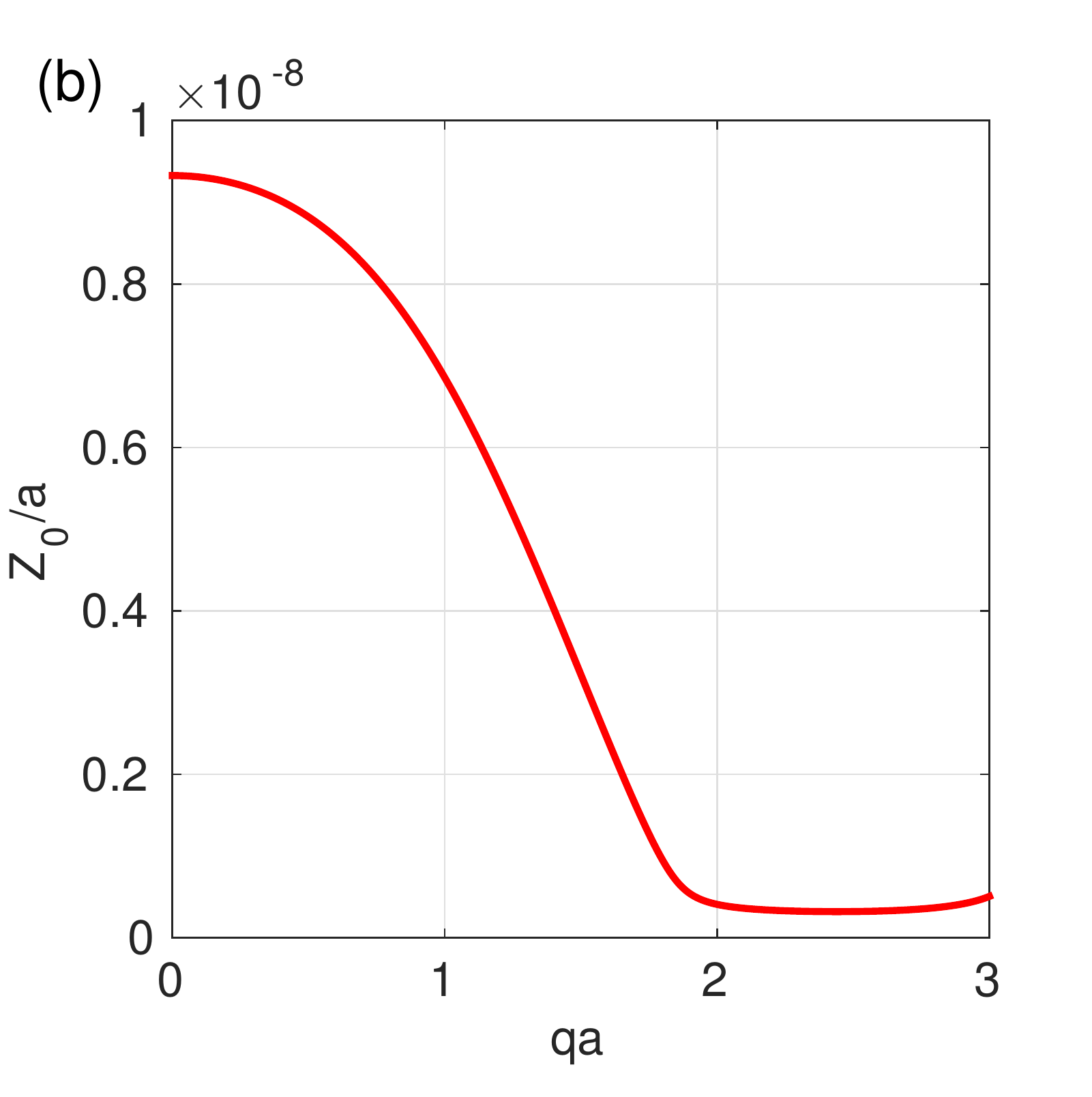}
\caption{The relative zero-point fluctuation, $Z_q/a$ vs.  $qa$, (a) for
  acoustic modes, and (b) for vibrational modes.}
\label{ZPF1}
\end{figure}

For illustration we consider a nanofiber of radius $a=250$~nm and length $L=1$~cm made of silicon material (density $\rho=2328$~Kg/m$^3$). In Figures~\ref{ZPF1}.a and \ref{ZPF1}.b we plot the zero-point-fluctuation relative to the fiber radius, $Z_q/a$,
as a function of $qa$, for the lowest two branches. It appears that the zero-point fluctuations decrease
with increasing $qa$ for the acoustic modes, where a singularity appears around $qa\approx 1.8$, and increase afterward for larger $qa$. The singularity
appears at the anti-crossing point among the acoustic and the lowest vibrational
modes, as seen in Figure~\ref{PhonDis}. The vibrational
modes have finite zero-point-fluctuations that decrease with increasing $qa$.

Importantly, the phonon frequencies for both acoustic and vibrational modes are in the GHz regime which allows to achieve low thermal occupation numbers at cryogenic temperatures. Moreover, mechanical quality factors measured in recent experiments with tapered nanofibres were in the range of $10^4$ \cite{Wuttke2013}. For square-shaped waveguides mechanical quality factors of several 100 have been measured \cite{VanLaer2015b,Kittlaus2015}. In view of the tremendous progress made regarding mechanical quality factors of single-mode optomechanical systems we expect that there is vast room for improvement here with optimized nanostructures.

\subsection{Photon-Phonon Interactions}

Inserting the analytical expressions for the nanowire photon and phonon modes derived in the previous sections into the general expression for the photon-phonon interaction Hamiltonian \eqref{eq:HintQ} yields
\begin{equation}\label{eq:intH}
\hat{H}_{I}=\hbar\sum_{kq}\left\{f^{\ast}_{kq}~\hat{b}_q^{\dagger}\hat{a}_{k-q}^{\dagger}\hat{a}_k+f_{kq}~\hat{a}_k^{\dagger}\hat{a}_{k-q}\hat{b}_q\right\},
\end{equation}
where we exploited the translational symmetry along the waveguide axis implying
\begin{equation}
\frac{1}{L}\int_0^L \mathrm{d}ze^{-i(k-q-k')z}=\delta_{k',k-q}.
\end{equation}
Translational symmetry
results in conservation of
momentum in which two photons of wavenumbers $k$ and $k-q$ scatter by emission
or absorption of a phonon of wavenumber $q$.
\begin{figure}
\includegraphics[width=0.6\columnwidth]{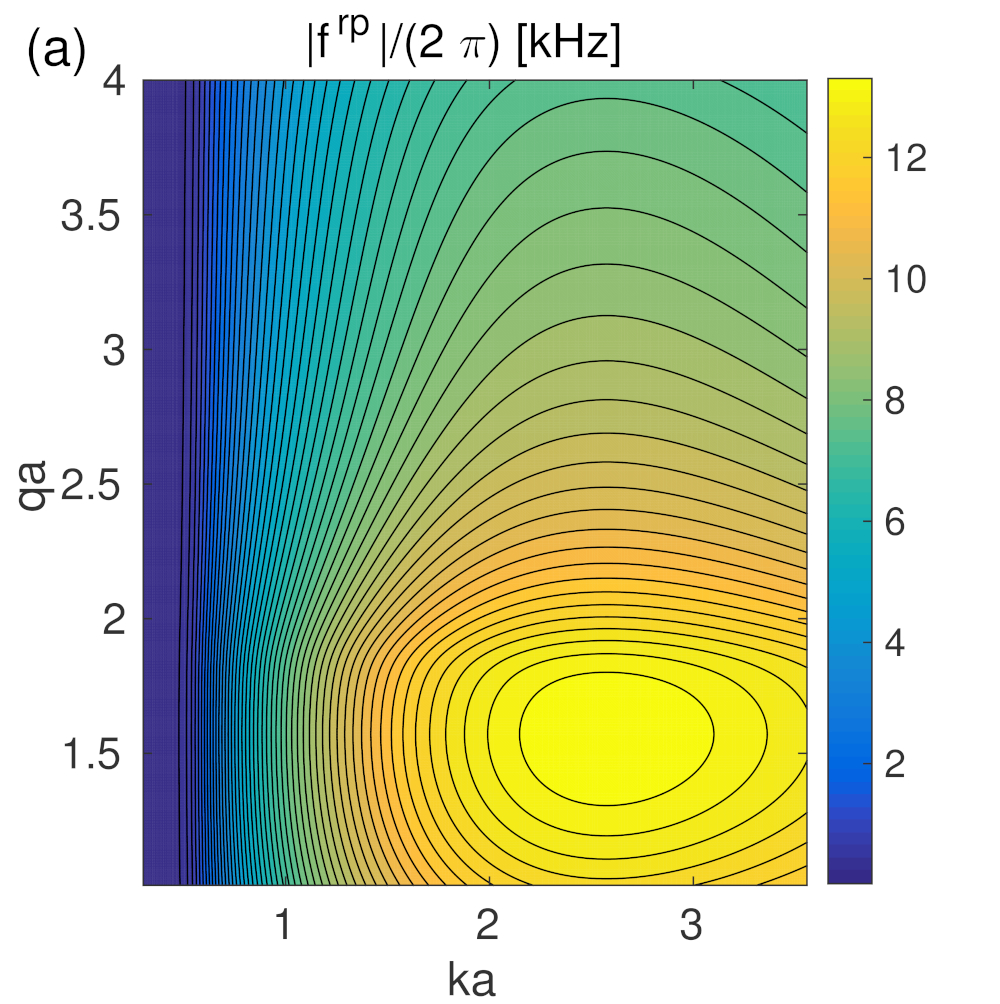}
\includegraphics[width=0.6\columnwidth]{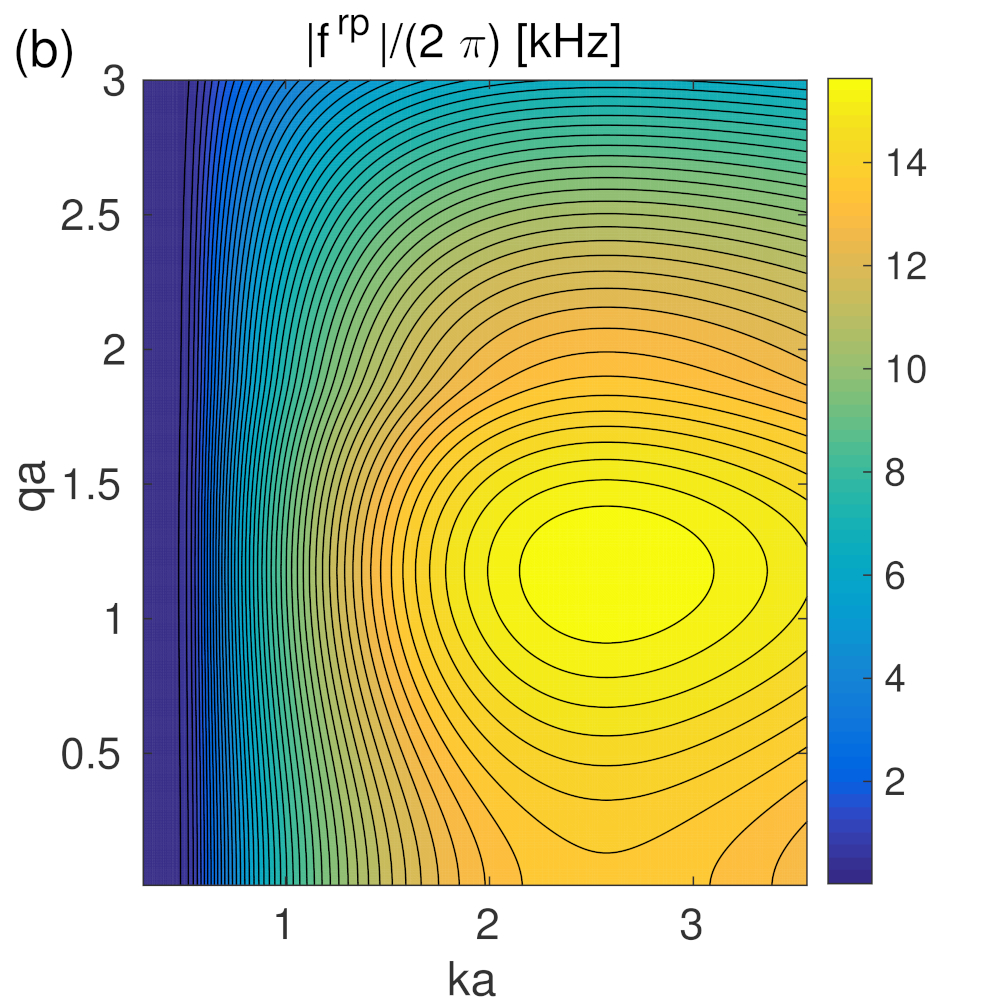}
\caption{The radiation pressure coupling parameter $|f_{kq}^\mathrm{rp}|/(2\pi)$
  contour vs. the plane $(ka-qa)$, for
  scattering that involves: (a) acoustic modes, and (b) vibrational modes. The
  fiber length is $L=1$~cm. The optomechanical parameters $f_{kq}^\mathrm{rp}$ scale as $L^{-1/2}$. }
\label{Coup1}
\end{figure}
The coupling is $f_{kq}=f_{kq}^\mathrm{rp}+f_{kq}^\mathrm{el}$, as in Eq.~\eqref{eq:HintQ}. The coupling parameter due  to radiation pressure is given by
\begin{equation} \label{RPP}
f_{kq}^\mathrm{rp}=-\frac{Z_q}{a_k^\mathrm{eff}}\sqrt{\omega_k\omega_{k-q}}F_{kq}^\mathrm{rp},
\end{equation}
where
\begin{eqnarray}
F_{kq}^\mathrm{rp}&=&\frac{a}{a_k^\mathrm{eff}}\frac{n^2-1}{2}w^r_q\nonumber \\
&\times&\left\{u_k^{z\ast<}u_{k-q}^{z<}+u_k^{\theta\ast<}u_{k-q}^{\theta<}+n^2u_k^{r\ast<}u_{k-q}^{r<}\right\},
\end{eqnarray}
and all vector mode functions are evaluated on the fiber surface according to
Eqs.~\eqref{eq:u_inner}. The effective radius of the  photon mode
$a_k^\mathrm{eff}$ is defined through $V_k^\mathrm{phot}=\pi
(a_k^\mathrm{eff})^2L$. In view of $\omega_q\ll\omega_k$ we approximate
$\omega_{k-q}\simeq\omega_k$ in the following. The dependence of $|f_{kq}|$ on
$L$ is only through $Z_q$ that gives $|f_{kq}|\propto 1/\sqrt{L}$.

In Figs.~\ref{Coup1}.a and \ref{Coup1}.b we show the radiation pressure
coupling parameter, $|f_{kq}^\mathrm{rp}|/(2\pi)$, versus $ka$ and $qa$ for
scattering that involves, respectively, the acoustic modes and the lowest
vibrational modes for a fiber of length $L=1$~cm.  Radiation pressure coupling parameters have high values of about $10$~kHz, in the region of $qa\approx 1.5-2$, for acoustic modes and $qa\approx 1-1.5$ for vibrational modes, which appear in the region of $ka\approx 2-3$ for photons. This region for optical photons appears at nanoscale waveguides, and the coupling is significantly decreased at microscale and larger structures.

\begin{figure}
\includegraphics[width=0.6\columnwidth]{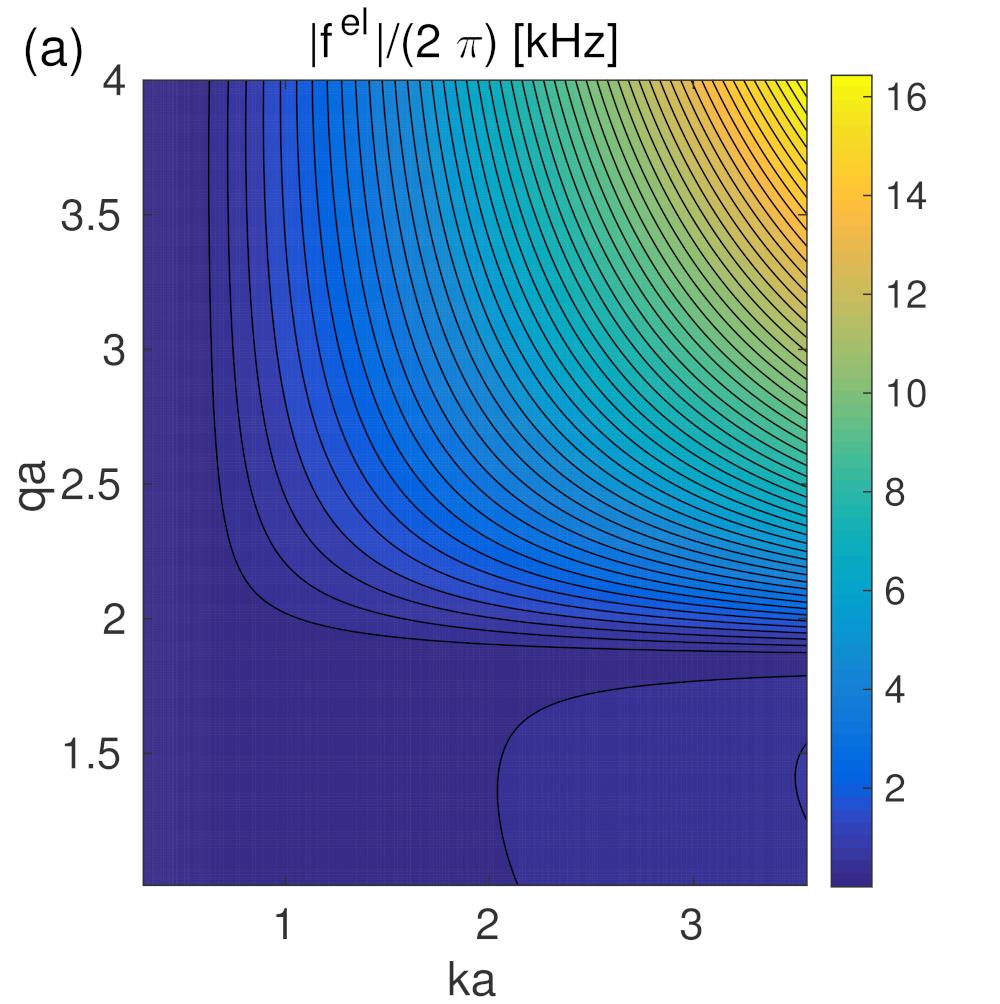}
\includegraphics[width=0.6\columnwidth]{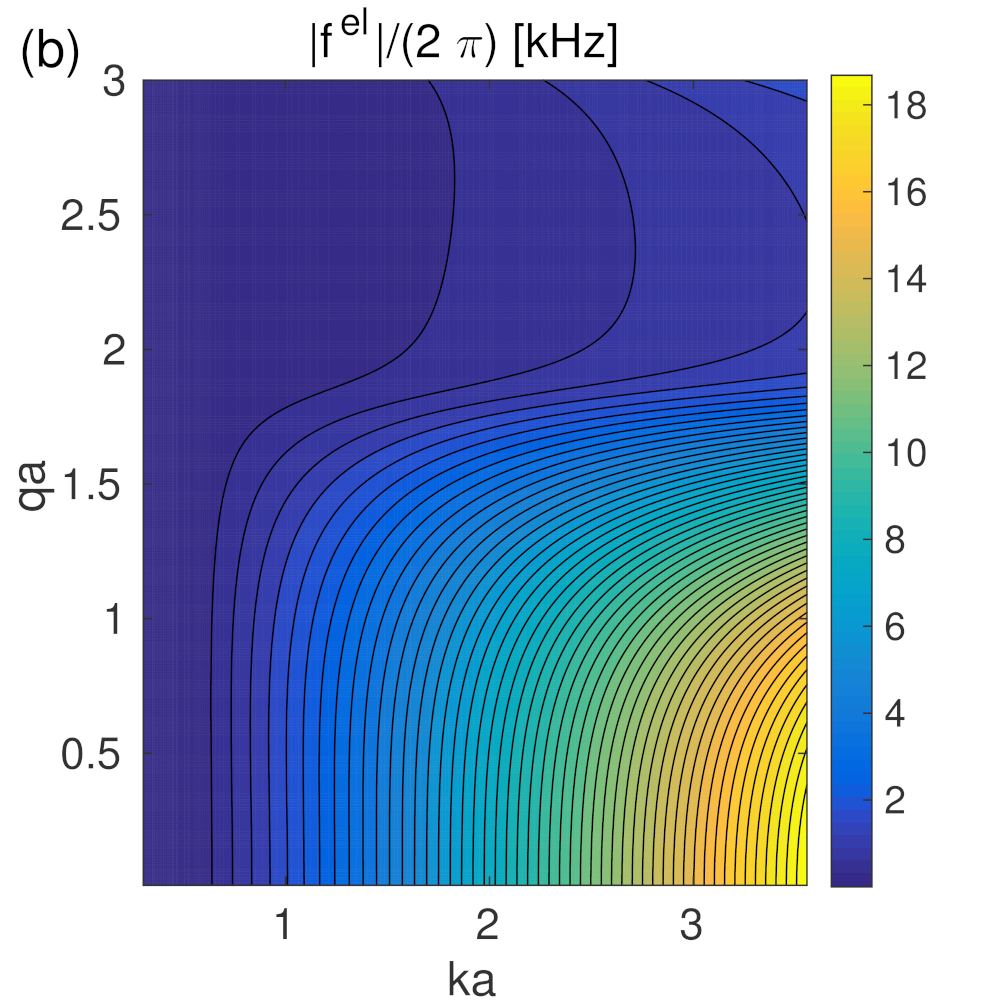}
\caption{The electrostriction coupling parameter $|f_{kq}^\mathrm{el}|/(2\pi)$
  contour vs. the plane $(ka-qa)$, for the
  scattering that involves: (a) acoustic modes, and (b) vibrational modes. The
  fiber length is $L=1$~cm. The optomechanical parameters $f_{kq}^\mathrm{el}$ scale as $L^{-1/2}$}
\label{CoupEl1}
\end{figure}

The coupling parameter due to electrostriction is given by
\begin{equation} \label{ELP}
f_{kq}^\mathrm{el}=\frac{Z_q}{a_k^\mathrm{eff}}{\sqrt{\omega_k\omega_{k-q}}}F_{kq}^\mathrm{el},
\end{equation}
where
\begin{eqnarray}
F_{kq}^\mathrm{el}&=&\frac{\gamma_\mathrm{el}}{2a_k^\mathrm{eff}}\int_0^ardr\left({\bf
  u}_k^{<\ast}(r)\cdot{\bf u}^<_{k-q}(r)\right)\nonumber \\
&\times&\left\{\frac{1}{r}\frac{\partial}{\partial
  r}\left(rw^r_q(r)\right)+iqw^z_q(r)\right\}.
\end{eqnarray}
In Figs.~\ref{CoupEl1}a and \ref{CoupEl1}b we show the coupling parameter $|f_{kq}^\mathrm{el}|/(2\pi)$ versus $ka$ and $qa$, for
scattering involving, respectively, the acoustic modes and the lowest
vibrational modes again for a fiber of length $L=1$~cm. The electrostriction
parameter for dielectric materials can be written as $\gamma_\mathrm{el}\approx n^4p_{12}$, where
$p_{12}$ is the elasto-optic parameter. For silicon we have $n\approx 3.5$
and $p_{12}\approx 0.017$, hence we get $\gamma_\mathrm{el}\approx 2.55$.  For acoustic modes the electrostriction coupling parameters
increase with increasing $ka$ and $qa$, while for vibrational modes they increase with increasing $ka$ only at small $qa$.

The comparison between the two coupling mechanisms show that electrostriction is small in zones
where radiation pressure is maximal, which is the case for optical light in
nanoscale waveguides. But for microscale and larger waveguides
electrostriction becomes dominant where radiation pressure is significantly
suppressed. Nanowires of square cross
sections are expected to give larger photon-phonon coupling parameters due to
the fact that light is more concentrated on the boundary \cite{Rakish2012}, but
they have smaller mechanical quality factor relative to cylindrical nanofibers.

\section{Real-Space Representation and Brillouin Gain Parameter}\label{sec:Gain}

In this section we transform the coupled photon-phonon Hamiltonian from momentum-space to
real-space representation. This is especially instructive in cases when the description can be effectively constrained to relatively narrow frequency bands, as is the case when narrowband light is injected into the nanofibre and at the same time Brillouin scattering populated only selected narrow bands of phonons. The real-space representation developed in the following provides coupled one dimensional propagation equations for narrowband photons and phonons. As a first application we derive the Brillouin gain parameters for nanofibres measured in recent experiments from our ab initio calculation of the optomechanical coupling parameters $f_{kq}^\mathrm{rp}$ and $f_{kq}^\mathrm{el}$.

\subsection{Real-Space Representation}

For the effectively one-dimensional photon field introduced in Sec.~\ref{sec:phot} we define associated operators in real space as
\begin{align*}
  \hat\psi(z)&=\frac{1}{\sqrt{L}}\sum_{k\in B_{k_0}}\hat{a}_k e^{i(k-k_0)z}.
\end{align*}
Here $B_{k_0}$ denotes a suitable bandwidth of photon wave numbers centered around a central wave numbers $k_0$. The operator $\hat\psi(z)$ defined such as to describe slowly varying spatial amplitudes relative to the wave $e^{ik_0z}$. For positive (negative) sign of $k_0$ the slowly varying operators $\hat\psi(z)$ describe right (left) propagating photons. The definition of $\hat\psi(z)$ implies $[\hat\psi(z),\hat\psi^\dagger(z')]=i\delta(z-z')$ where the $\delta$-function is understood to be of width $\sim B^{-1}_{k_0}$. The inverse relation is
\begin{align*}
  \hat{a}_k&=\frac{1}{\sqrt{L}}\int_0^L \mathrm{d}z\,\hat{\psi}(z)e^{-i(k-k_0)z}.
\end{align*}
Furthermore, we approximate the photon dispersion relation shown in Fig.~\ref{PhotDis} within the relevant bandwidth $B_{k_0}$ as $\omega_k=\omega_{k_0}+v_g(k-k_0)$,
where $\omega_{k_0}$ is the bandwidth's central frequency, and $v_g=\partial\omega_k/\partial k$ is the group velocity. For the silicon nanowire considered here we have $v_g\simeq c/5$ in the range of wave numbers $1\lesssim ka\lesssim 3$, cf. Fig.~\ref{PhotDis}. In real space representation the Hamiltonian \eqref{eq:Hphot} of the free photon field for modes within the bandwidth $B_{k_0}$ is
\begin{equation}
\hat{H}_\mathrm{phot}=\hbar\omega_{k_0}\int \mathrm{d}z~\hat{\psi}^{\dagger}(z)\hat{\psi}(z)
 -i\hbar v_g\int \mathrm{d}z~\hat{\psi}^{\dagger}(z)\frac{\partial\hat{\psi}(z)}{\partial z}.
\end{equation}

In complete analogy, we define for the one dimensional fields in Sec.~\ref{sec:phon} of acoustic phonons and vibrational phonons  the real space operators
\begin{align*}
\hat Q(z)&=\frac{1}{\sqrt{L}}\sum_{q\in B_{q_0}}\hat{b}_q e^{i(q-q_0)z},
\end{align*}
with inverse relation
\begin{equation}
\hat{b}_q=\frac{1}{\sqrt{L}}\int \mathrm{d}z~\hat{Q}(z)e^{-i(q-q_0)z}.
\end{equation}
and commutation relation $[\hat{Q}(z),\hat{Q}^\dagger(z')]=\delta(z-z')$.  The phonons have a linear dispersion with sound velocity $v_s$, such that within the bandwidth  $B_{q_0}$ we approximate $\Omega_q=\Omega_{q_0}+v_s(q-q_0)$. Thus, the free Hamiltonian \eqref{eq:phonH} for phonons is
\begin{equation}
\hat{H}_\mathrm{phon}=\hbar\Omega_{q_0}\int \mathrm{d}z~\hat{Q}^{\dagger}(z)\hat{Q}(z)-i\hbar v_s\int
\mathrm{d}z~\hat{Q}^{\dagger}(z)\frac{\partial\hat{Q}(z)}{\partial z}.
\end{equation}
The vibrational modes are almost dispersion-less below $qa\simeq 2$ such that $v_s\simeq 0$ in this regime.

Finally, the interaction Hamiltonian \eqref{eq:intH} is given by
\begin{eqnarray}
\hat{H}_{I}&=&\sqrt{L}\hbar\int \mathrm{d}z\nonumber \\
&\times&\left\{f^{\ast}~\hat{Q}^{\dagger}(z)\hat{\psi}^{\dagger}(z)\hat{\psi}(z)+f~\hat{\psi}^{\dagger}(z)\hat{\psi}(z)\hat{Q}(z)\right\},
\end{eqnarray}
where $f$ is the photon-phonon coupling parameter in the local field
approximation, in which we neglect the weak dependence of $f$ on wavenumbers within the relevant bandwidths $B_{q_0}$ and $B_{k_0}$ of phonon and photon modes.

\begin{figure}
\includegraphics[width=0.6\columnwidth]{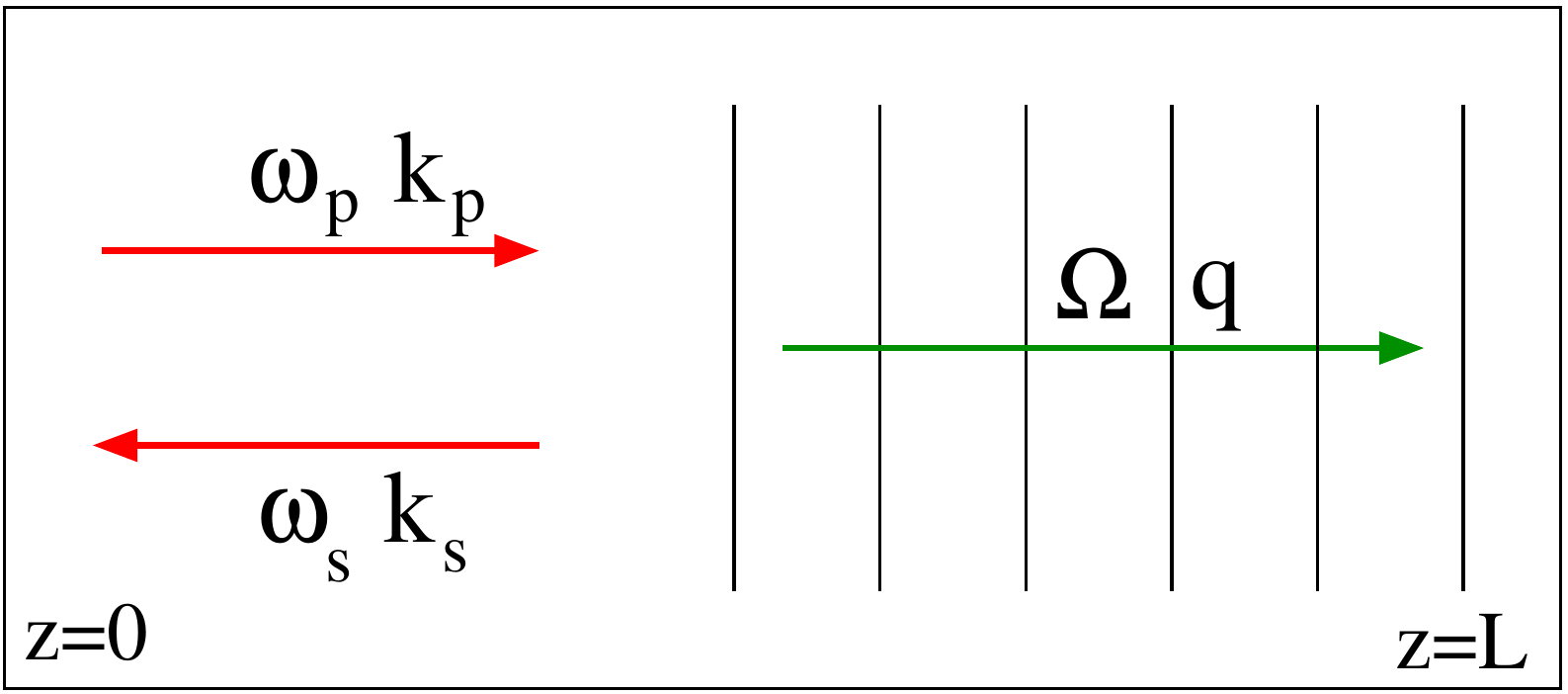}
\caption{The backward Stokes SBS. The pump field of frequency $\omega_p$ and
  wave number $k_p$ is scattered into the Stokes field  of frequency $\omega_s$ and
  wave number $k_s$, and a sound wave of frequency $\Omega$ and wave number
  $q$.}
\label{SBS}
\end{figure}

\subsection{Brillouin Gain Parameter}

\begin{figure}
\includegraphics[width=0.6\columnwidth]{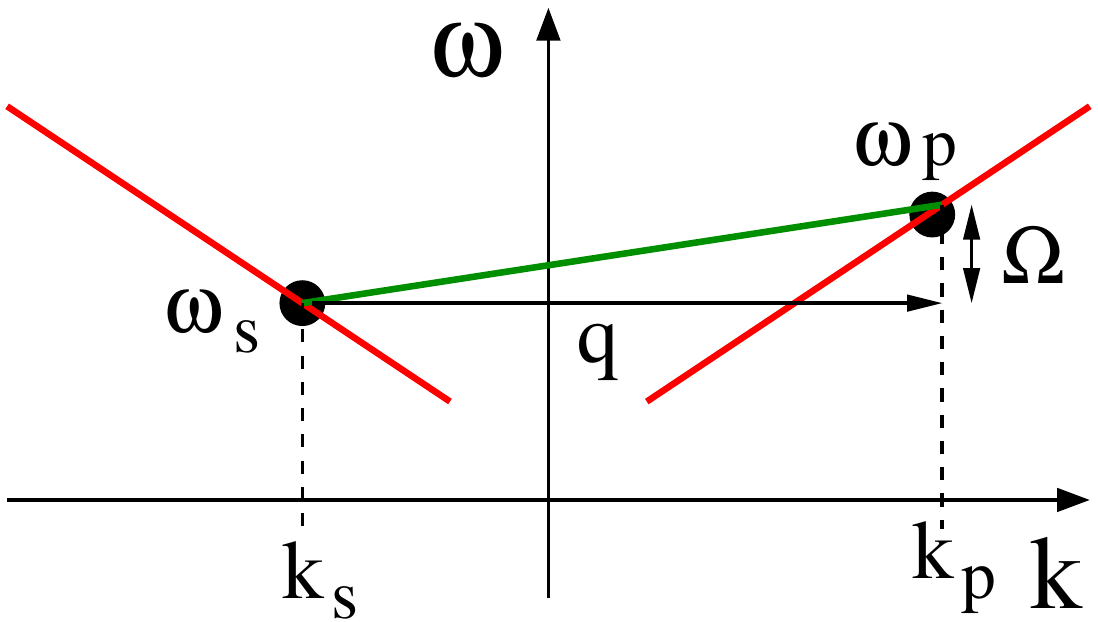}
\caption{The backward Stokes SBS. The process is presented schematically on the dispersion plot in
  order to emphasize the conservation of energy $\omega_p=\omega_s+\Omega$ and momentum $q=k_s+k_p$.}
\label{PumpStokes}
\end{figure}

The real space description from the previous section generalizes immediately when more than one band of phonon or photon modes are considered, as we will discuss now for the case of stimulated Brillouin scattering. The result of this treatment will be a direct relation between the optomechanical coupling parameters $f_{kq}$ and the Brillouin gain parameter which can be observed directly in experiments. We consider the backward Brillouin scattering among two narrow band light fields, $(s)$ the Stokes probe field and $(p)$ the strong pump field, involving acoustic phonons, as seen in Fig.~\ref{SBS}. We denote the central frequencies for the three bands by $\omega_s$, $\omega_p$ and $\Omega$, respectively, and assume energy conservation, $\omega_p=\omega_s+\Omega$, and momentum conservation,  ${\bf k}_p={\bf k}_s+{\bf q}$. These conditions are illustrated in Fig.~\ref{PumpStokes}.

Generalizing the results from the previous section to this configuration, the real-space Hamiltonian is
\begin{align}\label{eq:SBS}
\hat{H}&=-i\hbar v_s\int
\mathrm{d}z~\hat{Q}^{\dagger}(z)\frac{\partial\hat{Q}(z)}{\partial
  z}-i\hbar v_g\int
\mathrm{d}z~\hat{\psi}^{\dagger}_{p}(z)\frac{\partial\hat{\psi}_{p}(z)}{\partial
  z}\nonumber\\
&+i\hbar v_g\int
\mathrm{d}z~\hat{\psi}^{\dagger}_{s}(z)\frac{\partial\hat{\psi}_{s}(z)}{\partial
  z}+\sqrt{L}\hbar\int
\mathrm{d}z\nonumber\\
&\times\left\{f^{\ast}~\hat{Q}^{\dagger}(z)\hat{\psi}_s^{\dagger}(z)\hat{\psi}_p(z)+f~\hat{\psi}_p^{\dagger}(z)\hat{\psi}_s(z)\hat{Q}(z)\right\},
\end{align}
This Hamiltonian is written in an interaction picture with respect to
\begin{align}
\hat{H}_0&=\hbar\int\mathrm{d}z\left(\omega_p~\hat{\psi}_p^{\dagger}(z)\hat{\psi}_p(z)
+\omega_s~\hat{\psi}_s^{\dagger}(z)\hat{\psi}_s(z)\right.\nonumber\\
&+\left.\Omega~\hat{Q}^{\dagger}(z)\hat{Q}(z)\right),
\end{align}
and under a rotating wave approximation where all non-resonant terms (such as $\hat{Q}^{\dagger}\hat{\psi}_p^{\dagger}\hat{\psi}_s$ or $\hat{Q}^{\dagger}\hat{\psi}_p^{\dagger}\hat{\psi}_p$) were dropped.

The equations of motion corresponding to the Hamiltonian in Eq.~\eqref{eq:SBS} are
\begin{align}
\left(\frac{\partial}{\partial t}+v_g\frac{\partial}{\partial
  z}\right){\hat{\psi}}_{p}(z,t)&=-i\sqrt{L}f{\hat{Q}}(z,t){\hat{\psi}}_s(z,t), \nonumber \\
\left(\frac{\partial}{\partial t}-v_g\frac{\partial}{\partial
  z}\right){\hat{\psi}}_{s}(z,t)&=-i\sqrt{L}f^{\ast}{\hat{Q}}^{\dagger}(z,t){\hat{\psi}}_p(z,t), \nonumber \\
\left(\frac{\partial}{\partial t}+v_s\frac{\partial}{\partial
  z}\right){\hat{Q}}(z,t)&=-\frac{\Gamma}{2}{\hat{Q}}(z,t) \nonumber \\
&-i\sqrt{L}f^{\ast}{\hat{\psi}}_s^{\dagger}(z,t){\hat{\psi}}_p(z,t)+\hat{F}(z,t),
\end{align}
where $\Gamma$ is the acoustic phonon damping rate, and we neglect the
photon damping. $\hat{F}(z,t)$ is the Langevin noise operator. The damping
rate parameter can be extracted from the observed mechanical quality factor
$Q$ and in
knowing the phonon frequency \cite{VanLaer2015,Rakish2012,Shin2013,VanLaer2015a,VanLaer2015b,Kittlaus2015}. At steady state all time derivatives for the slowly varying operators vanish and, moreover, $\frac{\partial}{\partial   z}{\hat{Q}}(z,t)$ can be neglected for acoustic phonons. Hence we get
\begin{equation}
{\hat{Q}}(z,t)\approx -i\frac{2\sqrt{L}f^{\ast}}{\Gamma}{\hat{\psi}}_s^{\dagger}(z,t){\hat{\psi}}_p(z,t)+\frac{2}{\Gamma}\hat{F}(z,t).
\end{equation}
Substitutions of ${\hat{Q}}(z,t)$ in the equations of motion for the photon fields yields
\begin{align}
\frac{\partial}{\partial
  z}{\hat{\psi}}_{p}(z,t)&=-\frac{2L|f|^2}{v_g\Gamma}{\hat{\psi}}_s^{\dagger}(z,t){\hat{\psi}}_s(z,t){\hat{\psi}}_p(z,t) \nonumber \\
&-i\frac{2\sqrt{L}f}{\Gamma v_g}\hat{F}(z,t){\hat{\psi}}_{s}(z,t),
\nonumber \\
\frac{\partial}{\partial
  z}{\hat{\psi}}_{s}(z,t)&=-\frac{2L|f^{a}|^2}{v_g\Gamma}{\hat{\psi}}_p^{\dagger}(z,t){\hat{\psi}}_p(z,t){\hat{\psi}}_s(z,t) \nonumber \\
&-i\frac{2\sqrt{L}f^{\ast}}{\Gamma v_g}\hat{F}^{\dagger}(z,t){\hat{\psi}}_{p}(z,t).
\end{align}
The light field intensity is defined by \cite{Loudon2000}
\begin{align}
I_{p}&=v_g\frac{\hbar\omega}{\cal   A}\langle{\hat{\psi}}_p^{\dagger}(z,t){\hat{\psi}}_p(z,t)\rangle,\\
I_s&=v_g\frac{\hbar\omega}{\cal   A}\langle{\hat{\psi}}_s^{\dagger}(z,t){\hat{\psi}}_s(z,t)\rangle,
\end{align}
where we set $\omega=\omega_s\simeq\omega_p$, and denote by ${\cal A}$ the
waveguide cross section. The strong pump field is considered as a classical
field, and we use the Langevin force properties $\langle\hat{F}(z,t){\hat{\psi}}_{s}(z,t)\rangle=\langle\hat{F}^{\dagger}(z,t){\hat{\psi}}_{s}^{\dagger}(z,t)\rangle=0$. Then we have
\begin{align}\label{eq:SBSeoms}
\frac{\partial}{\partial   z}I_s&=-G_B\mathcal{A}I_pI_s, \\
\frac{\partial}{\partial   z}I_p&=-G_B\mathcal{A}I_pI_s,
\end{align}
where the Brillouin gain factor is defined by
\begin{equation}
G_B=\frac{4L|f|^2}{\hbar\omega v_g^2\Gamma}.
\end{equation}
Neglecting the pump depletion, then the first of Eqs.~\eqref{eq:SBSeoms} can be integrated to give $\hat{I}_s(0)=\hat{I}_s(L)\exp(G_BI_pV)$. Note that the Stokes field is propagating to the left, such that $\hat{I}_s(0)$ described the outgoing intensity. Thus, the gain parameter $G_B$ expresses the gain in Stokes intensity per medium volume and pump intensity. We remark also that the coupling parameters in Eqs. \eqref{RPP}) and \eqref{ELP} are proportional to  $L^{-1/2}$ which makes $|f|^2L$ and therefore also $G_B$ independent of $L$.

We consider the scattering of the photons maximally localized in transverse direction with $ka\approx 1.74$ that is, for $a=\unit[250]{nm}$, of wave length $\lambda\simeq 900$~nm. The phonons for the backward scattering are of $qa\approx
3.48$ with the damping rate of $\Gamma/(2\pi)=\unit[1.5]{MHz}$. The photon-phonon
coupling parameter is $f/(2\pi)\approx \unit[5]{kHz}$, for a nano-waveguide of length $L=\unit[1]{cm}$. For $\Omega/(2\pi)=\unit[15]{GHz}$ the quality
factor is $Q=10^4$. The group velocity is about
$v_g\approx c/5$. The gain efficiency
is about ${G}_B\approx \unit[10^4]{m^{-1}W^{-1}}$, which agrees with the
experimental results \cite{VanLaer2015b}. The values of all parameters used
here are summarized in table \eqref{table}.

The result provides a relation between the observable gain parameter $G_B$ and the
photon-phonon coupling parameter $f$, which agree with the one derived in \cite{VanLaer2015c}. The relations allow us to compare the
value of the calculated photon-phonon coupling parameter to the experimental
value of the gain parameter. A similar result holds for the forward Brillouin scattering
involving vibrational modes.

\begin{table}[ht]
\caption{Values for case study on silicon nanowire}
\centering
\begin{tabular}{c c c}
\hline\hline
Name & Symbol & Value \\ [0.5ex]
\hline
Fiber radius & $a$ & $250$~nm \\
Fiber length & $L$ & $1$~cm \\
Photon group velocity & $v_g$ & $c/5$ \\
Photon wavelength & $\lambda$ & $900$~nm $(ka\approx 1.74)$ \\
Acoustic phonon frequency & $\Omega/(2\pi)$ & $15$~GHz $(qa\approx 3.48)$ \\
Phonon damping rate & $\Gamma/(2\pi)$ & $1.5$~MHz \\
Mechanical quality factor & $Q$ & $10^4$ \\
Photon-phonon coupling & $f/(2\pi)$ & $5$~kHz \\
Gain parameter & $G_B$ & $10^4$~m$^{-1}$W$^{-1}$ \\
\hline
\end{tabular}
\label{table}
\end{table}

\section{Summary}

Starting from the classical electromagnetic field Hamiltonian in dielectric
media we derived the mutual coupling between light and mechanical
excitations. The interaction Hamiltonian is obtained by perturbing the medium
due to mechanical excitations. We treated two type of fluctuations in the dielectric medium. The
first is the fluctuation in the dielectric constant that generates
electrostriction coupling, and the second is fluctuations in the dielectric
material boundaries that introduces radiation pressure. In order to overcome
the difficulty due to the field jumps on the boundaries we expressed the
Hamiltonian in terms of continuous fields. The main objective in both cases is
to represent the fluctuations in terms of the mechanical displacement
vector. The derived Hamiltonian can be adopted for any structure of dielectric
medium ranging from nano resonators up to bulk materials.

In quantum mechanics the electromagnetic field is represented as photons and the
mechanical excitations as phonons, where the classical formulation allows
direct quantization by converting the classical electric field and the
displacement field into operators. We treated in much details the case of a cylindrical nanophotonic waveguide, in which photons and phonons can propagate along the
waveguide with continuum wavenumbers, and are strongly confined in the
transverse direction with discrete modes. We explicitly solved for the photon
and phonon dispersions and mode functions, where each discrete mode gives rise
to photon or phonon branch. We derived the multi-mode photon-phonon
interaction Hamiltonian for both electrostriction and radiation pressure
couplings. The lowest photon branch is for HE$_{11}$ photons, in which a single
photon mode can propagate within the fiber with linear dispersion of group
velocity $\sim c/5$, where for a silicon nanowire this mode is maximally localized
inside the waveguide around $ka\approx 1.74$ with a small part that penetrates
into the surrounding environment. The lowest phonon branch is of acoustic modes
with linear dispersion, and the lowest excited phonon branch is dispersion-less
of localized vibrational modes with frequency of about $\unit[10]{GHz}$. These two
phonon branches approach different linear dispersions beyond the anti-crossing
point.

We calculated the photon-phonon coupling parameters due to electrostriction
and radiation pressure mechanisms for a silicon nanofiber. For optical light
we found that radiation pressure dominates electrostriction coupling at
nanoscale waveguides with coupling parameter of about $\unit[10]{kHz}$, while electrostriction
becomes dominant at larger dimensions. We provide a tool for checking the
validity of the theoretically predicted coupling parameters by relating them
to the experimentally observed gain parameter. We consider Stokes backward
Brillouin scattering, where a strong pump field scatters into a Stokes field and a sound
wave. Starting from the real space representation of the photon and phonon
Hamiltonians, we solve the field equations of motion at steady state to get
the Stokes field amplification where the gain parameter is expressed in terms
of the photon-phonon coupling parameter.

The results presented here provide a general outline for deriving quantum optical multi-mode Hamiltonians for interacting photons and phonons in optomechanical nanophotonic structures. In view of the success achieved in recent years with optomechanical structures involving single phonon and photon modes we envision that their multimode counterparts offer great possibilities to observe and exploit quantum effects in extended nanophotonic media. In particular, the optomechanical Kerr-nonlinearity mediated by phonons can be exploited for the
study of quantum nonlinear optics and for many-body physics of strongly
correlated photons. Moreover, Brillouin induced transparency with the possibility of slow light, in analogy to EIT in cold dense atomic gases, can be achieved in the present nanoscale waveguides. The quantum optical Hamiltonian derived in the present work provides firm grounds for future explorations into these directions.

\section*{Acknowledgment}

We thank Raphael van Laer for useful comments. This work was
funded by the European Commission (FP7-Programme) through iQUOEMS (Grant
Agreement No. 323924). We acknowledge support by DFG through QUEST.

\appendix

\section{Perturbation Theory of Maxwell's Equations}\label{sec:Maxwell}

We will show here that Eq.~\eqref{deltaEMH} provides the relevant part of the correction to the field Hamiltonian for a given change of the permittivity, and determine the order of magnitude of the correction due to induced changes in the field amplitudes.  The equations of motion following from the field Hamiltonian \eqref{EMH} are Maxwell's equations
\begin{subequations}
\begin{align}\label{eq:Maxwell}
\nabla\times{\bf E}&=-\mu_0\frac{\partial{\bf H}}{\partial
  t},\\
  \nabla\cdot\left(\epsilon({\bf x}){\bf E}\right)&=0, \\
\nabla\times{\bf H}&=\epsilon({\bf x})\frac{\partial{\bf E}}{\partial
  t},\label{eq:H}\\
  \nabla\cdot{\bf H}&=0,
\end{align}
\end{subequations}
where ${\bf E}$ and ${\bf H}={\bf B}/\mu_0$ are the electric and magnetic fields, and $\epsilon({\bf x})$ is the (dimensionless) permittivity of the assumed dielectric, lossless, and nonmagnetic material. Maxwell's equations imply for the magnetic field
\[
\frac{1}{c^2}\frac{\partial^2 {\bf H} }{\partial t^2}+\nabla\times\left(\frac{\epsilon_0}{\epsilon({\bf x})}\nabla\times{\bf H}\right)=0.
\]
Harmonic solutions ${\bf H}({\bf x},t)={\bf H}_k({\bf x}) e^{-i\omega_k t}$ therefore fulfill the eigenvalue equation
\begin{align}\label{eq:evH}
\nabla\times\left(\frac{\epsilon_0}{\epsilon({\bf x})}\nabla\times{\bf H}_k\right)=\frac{\omega_k^2}{c^2}{\bf H}_k.
\end{align}
The operator on the left hand side is Hermitian \cite{Skorobogatiy2009} and, therefore, the eigenmodes are orthonormal,
$
\int_V\mathrm{d}V~{\bf H}^*_k({\bf x}){\bf H}_{l}({\bf x})=\delta_{kl}.
$
When the permittivity is changed slightly $\epsilon({\bf x})\rightarrow \epsilon({\bf x})+\delta\epsilon({\bf x})$ the perturbed eigenmodes ${\bf H}_k+\delta {\bf H}_k$ will remain orthonormal which implies for the first order corrections
\begin{align}\label{eq:deltaH}
\int_V\mathrm{d}V\left\{\delta{\bf H}^*_k({\bf x}){\bf H}_{l}({\bf x})+{\bf H}^*_k({\bf x})\delta{\bf H}_{l}({\bf x})\right\}=0.
\end{align}

On the other hand, the electric displacement field ${\bf D}=\epsilon({\bf x}){\bf E}$ corresponding  to field mode ${\bf H}_{k}$ is ${\bf D}_{k}=i\omega_k^{-1}\nabla\times{\bf H}_{k}$ due to Eq.~\eqref{eq:H}, and the same relation holds for the first order perturbation $\delta{\bf D}_{k}=i\omega_k^{-1}\nabla\times\delta{\bf H}_{k}$. Note that this statement does not apply in the same way to the electric field  ${\bf E}_k=i(\epsilon({\bf x})\omega_k)^{-1}\nabla\times{\bf H}_{k}$ whose perturbation contains, apart from $\delta{\bf H}_{k}$, further contributions proportional to $\delta\epsilon({\bf x})$. For a general electric displacement field ${\bf D}({\bf x},t)=\sum_k{\bf D}_{k}({\bf x})d_k(t)$ with amplitudes $d_k(t)$ one therefore finds
\begin{align*}
 &\int_V\mathrm{d}V \frac{1}{\epsilon({\bf x})}\delta{\bf D}^*{\bf D}\\
&=\sum_{k,l}\frac{d_k^*d_l}{\omega_k\omega_l}\int_V\mathrm{d}V\frac{1}{\epsilon({\bf x})}(\nabla\times\delta{\bf H}^*_k)(\nabla\times{\bf H}_l)\\
  &=\frac{1}{\epsilon_0}\sum_{k,l}\frac{d_k^*d_l}{\omega_k\omega_l}\int_V\mathrm{d}V\delta{\bf H}^*_k\nabla\times\left(\frac{\epsilon_0}{\epsilon({\bf x})}\nabla\times{\bf H}_l\right)\\
  &=\frac{1}{\epsilon_0}\sum_{k,l}\frac{d_k^*d_l}{c^2}\frac{\omega_k}{\omega_l}\int_V\mathrm{d}V\delta{\bf H}^*_k{\bf H}_l
\end{align*}
where we used the eigenvalue equation \eqref{eq:evH} in the last step. Overall, this implies
\begin{align*}
 & \int_V\mathrm{d}V \frac{1}{\epsilon({\bf x})}\left(\delta{\bf D}^*{\bf D}+{\bf D}^*\delta{\bf D}\right)\\
  &=\frac{1}{\epsilon_0}\sum_{k,l}\frac{d_k^*d_l}{c^2}\int_V\mathrm{d}V\left(\frac{\omega_k}{\omega_l}\delta{\bf H}^*_k{\bf H}_l
  +\frac{\omega_l}{\omega_k}{\bf H}^*_k\delta{\bf H}_l\right)\simeq 0,
\end{align*}
where we used Eq.~\eqref{eq:deltaH} and the fact that the optical photon frequency is not changed appreciably in Brillouin scattering, that is $\omega_k/\omega_l=1+\mathcal{O}(\omega_\mathrm{phonon}/\omega_\mathrm{photon})$. Thus, the  contribution to the perturbation of the field Hamiltonian due to the corrections in the field amplitudes, $\int_V\mathrm{d}V \epsilon({\bf x})^{-1}\delta|{\bf D}|^2$, is at most of magnitude $\mathcal{O}(\omega_\mathrm{phonon}/\omega_\mathrm{photon})$  to first order in $\delta\epsilon$ (or equivalently, in the mechanical displacement ${\bf Q}$). Note that this is not the case for $\int_V\mathrm{d}V \epsilon({\bf x})\delta|{\bf E}|^2$ for the reason given above. The reasoning presented here also shows that the first order correction of the Hamiltonian due to the perturbation of the magnetic field amplitudes vanishes exactly due to \eqref{eq:deltaH}.

\section{Nanofiber Electromagnetic Field}\label{App:FibreField}

Here we present the rigorous derivation of the nanofiber modes \cite{Jackson1999,Kien2004} for a cylindrical waveguide. We start from the Maxwell's equations \eqref{eq:Maxwell} in cylindrical
coordinates $(r,\theta,z)$. The fields $E_{\theta}$, $H_{\theta}$, $E_r$ and $H_r$ can be expressed in
terms of the axial fields $E_z$ and $H_z$ by
\begin{align} \label{EZTR}
E_r&=-i\frac{\beta}{\mu_0\epsilon\omega^2-\beta^2}\left(\frac{\partial}{\partial
  r}E_z+\frac{\mu_0\omega}{\beta}\frac{1}{r}\frac{\partial}{\partial\theta}H_z\right),
\nonumber \\
E_{\theta}&=-i\frac{\beta}{\mu_0\epsilon\omega^2-\beta^2}\left(\frac{1}{r}\frac{\partial}{\partial\theta}E_z-\frac{\mu_0\omega}{\beta}\frac{\partial}{\partial
  r}H_z\right), \nonumber \\
H_r&=-i\frac{\beta}{\mu_0\epsilon\omega^2-\beta^2}\left(\frac{\partial}{\partial
  r}H_z-\frac{\mu_0\omega}{\beta}\frac{1}{r}\frac{\partial}{\partial\theta}E_z\right),
\nonumber \\
H_{\theta}&=-i\frac{\beta}{\mu_0\epsilon\omega^2-\beta^2}\left(\frac{1}{r}\frac{\partial}{\partial\theta}H_z+\frac{\mu_0\omega}{\beta}\frac{\partial}{\partial
  r}E_z\right),
\end{align}
where $\beta$ is the axial wavenumber. The solution for the axial fields is given by
\begin{align}
E_z({\bf x},t)=R(r)e^{\pm il\theta}e^{-i(\beta z-\omega t)},  \nonumber \\
H_z({\bf x},t)=R(r)e^{\pm il\theta}e^{-i(\beta z-\omega t)},
\end{align}
where $l=0,1,2,\cdots$. The radial function $R(r)$ obeys the Bessel equation
\begin{equation}
\left\{\frac{1}{r}\frac{\partial}{\partial r}+\frac{\partial^2}{\partial r^2}+\left(k^2-\beta^2-\frac{l^2}{r^2}\right)\right\}R(r)=0,
\end{equation}
where $k^2=\mu_0\epsilon\omega^2$. The solution
inside the fiber, that is $r<a$, with $\epsilon=\epsilon_0n^2$, is given by
\begin{align}
E_z(r,\theta,z,t)&=AJ_l(pr)e^{-i(\beta z-\omega t\pm l\theta)}, \nonumber \\
H_z(r,\theta,z,t)&=BJ_l(pr)e^{-i(\beta z-\omega t\pm l\theta)},
\end{align}
where $p=\sqrt{n^2k_0^2-\beta^2}$, with $k_0=\omega/c$ is the free space wavenumber. The solution
on the outside of the fiber, that is $(r>a)$, with $\epsilon=\epsilon_0$, is given by
\begin{align}
E_z(r,\theta,z,t)&=CK_l(qr)e^{-i(\beta z-\omega t\pm l\theta)}, \nonumber \\
H_z(r,\theta,z,t)&=DK_l(qr)e^{-i(\beta z-\omega t\pm l\theta)},
\end{align}
where $q=\sqrt{\beta^2-k_0^2}$.

Using the relations (\ref{EZTR}), the solutions inside the fiber, that is $r<a$, are given by
\begin{align}
E_r(r,\theta,z,t)&=\left\{-i\frac{\beta}{p}AJ_l^{\prime}(pr)+\frac{\mu_0\omega}{p}(\pm
l)B\frac{J_l(pr)}{pr}\right\} \nonumber
\\
&e^{i(\omega t-\beta z\pm l\theta)}, \nonumber
\\
E_{\theta}(r,\theta,z,t)&=\left\{\frac{\beta}{p}(\pm
l)A\frac{J_l(pr)}{pr}+i\frac{\mu_0\omega}{p}BJ_l^{\prime}(pr)\right\} \nonumber
\\
&e^{i(\omega t-\beta z\pm l\theta)}, \nonumber
\\
E_z(r,\theta,z,t)&=AJ_l(pr)e^{i(\omega t-\beta z\pm l\theta)},
\end{align}
and
\begin{align}
H_r(r,\theta,z,t)&=\left\{-i\frac{\beta}{p}BJ_l^{\prime}(pr)-\frac{\epsilon\omega}{p}(\pm
l)A\frac{J_l(pr)}{pr}\right\} \nonumber
\\
&e^{i(\omega t-\beta z\pm l\theta)}, \nonumber
\\
H_{\theta}(r,\theta,z,t)&=\left\{\frac{\beta}{p}(\pm
l)B\frac{J_l(pr)}{pr}-i\frac{\epsilon\omega}{p}AJ_l^{\prime}(pr)\right\} \nonumber
\\
&e^{i(\omega t-\beta z\pm l\theta)}, \nonumber
\\
H_z(r,\theta,z,t)&=BJ_l(pr)e^{i(\omega t-\beta z\pm l\theta)}.
\end{align}
The solutions outside of the fiber, that is $r>a$, are given by
\begin{align}
E_r(r,\theta,z,t)&=\left\{i\frac{\beta}{q}CK_l^{\prime}(qr)-\frac{\mu_0\omega}{q}(\pm
l)D\frac{K_l(qr)}{qr}\right\} \nonumber
\\
&e^{i(\omega t-\beta z\pm l\theta)}, \nonumber
\\
E_{\theta}(r,\theta,z,t)&=\left\{-\frac{\beta}{q}(\pm
l)C\frac{K_l(qr)}{qr}-i\frac{\mu_0\omega}{q}DK_l^{\prime}(qr)\right\} \nonumber
\\
&e^{i(\omega t-\beta z\pm l\theta)}, \nonumber
\\
E_z(r,\theta,z,t)&=CK_l(qr)e^{i(\omega t-\beta z\pm l\theta)},
\end{align}
and
\begin{align}
H_r(r,\theta,z,t)&=\left\{i\frac{\beta}{q}DK_l^{\prime}(qr)+\frac{\epsilon_0\omega}{q}(\pm
l)C\frac{K_l(qr)}{qr}\right\} \nonumber
\\
&e^{i(\omega t-\beta z\pm l\theta)}, \nonumber
\\
H_{\theta}(r,\theta,z,t)&=\left\{-\frac{\beta}{q}(\pm
l)D\frac{K_l(qr)}{qr}+i\frac{\epsilon_0\omega}{q}CK_l^{\prime}(qr)\right\} \nonumber
\\
&e^{i(\omega t-\beta z\pm l\theta)}, \nonumber
\\
H_z(r,\theta,z,t)&=DK_l(qr)e^{i(\omega t-\beta z\pm l\theta)}.
\end{align}
The plus sign $(+l)$ is
refers to the right-handed solution of the transverse field, and the minus
sign $(-l)$ refers to the left-handed solution. The linearly polarized
solution can be composed as a superposition of right and left-handed solutions.

The constants $A$, $B$, $C$ and $D$ can be fixed by the boundary condition between
the inside and outside fields on the fiber surface, where the fields $E_z$,
$E_{\theta}$, $H_z$ and $H_{\theta}$ are continuous on the boundary. For the constants we obtain the relations
\begin{align}
C&=A\frac{J_l(pa)}{K_l(qa)},\ D=CB/A,\nonumber \\
B&=A\frac{i\beta(\pm
  l)}{\mu_0\omega}\left[\frac{1}{(pa)^2}+\frac{1}{(qa)^2}\right]\nonumber \\
&\times\left(\frac{J_l^{\prime}(pa)}{paJ_l(pa)}+\frac{K_l^{\prime}(qa)}{qaK_l(qa)}\right)^{-1},
\end{align}
and $A$ can be fixed by the normalization condition. The continuity leads to
a characteristic equation that determines the wavenumber $\beta$. For the EH modes, in which $E_z$ is larger
than $H_z$, we get
\begin{align}
\frac{J_{l-1}(pa)}{paJ_l(pa)}&=\left(\frac{1+n^2}{2n^2}\right)\frac{K_{l-1}(qa)+K_{l+1}(qa)}{2qaK_l(qa)}+\frac{l}{p^2a^2} \nonumber \\
\quad&+\left\{\left(\frac{n^2-1}{2n^2}\right)^2\left(\frac{K_{l-1}(qa)+K_{l+1}(qa)}{2qaK_l(qa)}\right)^2\right. \nonumber \\
&+\left.\left(\frac{l\beta}{nk_0}\right)^2\left(\frac{1}{q^2a^2}+\frac{1}{p^2a^2}\right)^2\right\}^{1/2},
\end{align}
and for the HE modes, in which $E_z$ is smaller
than $H_z$, we get
\begin{align}\label{Rot}
\frac{J_{l-1}(pa)}{paJ_l(pa)}&=\left(\frac{1+n^2}{2n^2}\right)\frac{K_{l-1}(qa)+K_{l+1}(qa)}{2qaK_l(qa)}+\frac{l}{p^2a^2} \nonumber \\
\quad&-\left\{\left(\frac{n^2-1}{2n^2}\right)^2\left(\frac{K_{l-1}(qa)+K_{l+1}(qa)}{2qaK_l(qa)}\right)^2\right. \nonumber \\
&+\left.\left(\frac{l\beta}{nk_0}\right)^2\left(\frac{1}{q^2a^2}+\frac{1}{p^2a^2}\right)^2\right\}^{1/2}.
\end{align}
The equations can be solved numerically or graphically and give a set of
discrete solutions for $\beta$ at each $l$ that are specified by index
$m$. The modes are labeled by HE$_{lm}$ and EH$_{lm}$. The transverse modes EH$_{0m}$ are
usually denoted by TM$_{0m}$, where $H_z$ vanishes. The transverse modes HE$_{0m}$ are
usually denoted by TE$_{0m}$, where $E_z$ vanishes. The modes HE$_{lm}$ and
EH$_{lm}$ are termed hybrid modes, as all field components are non-zero. The
hybrid modes represent screw rays and $l$ is associated with the orbital
angular momentum along the fiber axis.

\section{Nanofiber Mechanical Vibrations}\label{App:FibrePhonons}

We present the elastic waves in a cylindrical waveguide of isotropic medium
\cite{Achenbach1975}. The displacement vector is given by ${\bf Q}=(Q_r,Q_{\theta},Q_z)$. In the theory of elasticity the displacement can be derived
directly from the scalar $\phi$ and vector $\vec{\Psi}$ potentials. In general we can make the decomposition ${\bf Q}={\bf Q}_l+{\bf Q}_t$, where ${\bf Q}_l=\nabla\phi$ is a divergence-free vector, and
${\bf Q}_t=\nabla\times\vec{\Psi}$ is an irrotational vector. Hence, the material displacements are obtained from the
potentials by
\begin{equation}
{\bf Q}=\nabla\phi+\nabla\times\vec{\Psi}.
\end{equation}
The scalar field obeys the wave equation
\begin{equation}
\nabla^2\phi-\frac{1}{v_l^2}\frac{\partial^2\phi}{\partial t^2}=0,
\end{equation}
where $v_l$ is the velocity of the longitudinal wave, in which $\nabla\times{\bf Q}_l=0$. The vector potential obeys
the wave equation
\begin{equation}
\nabla^2\vec{\Psi}-\frac{1}{v_t^2}\frac{\partial^2\vec{\Psi}}{\partial t^2}=0,
\end{equation}
where $v_t$ is the velocity of the transverse wave, in which $\nabla\cdot{\bf
  Q}_t=0$. As the three displacements are expressed in terms of four scalar
potentials, one need an extra relation between the potentials. The
constrain condition usually used is $\nabla\cdot\vec{\Psi}=0$, but the
$\nabla\cdot\vec{\Psi}\neq0$ condition is also possible. The two displacement parts propagate
independently. The longitudinal component, ${\bf Q}_l$, propagates with velocity $v_l$, and the transverse component, ${\bf Q}_t$, propagates with velocity $v_t$.

Now we concentrate in the solution of the wave equations in cylindrical
coordinates. For harmonic waves of frequency $\Omega$, we have
\begin{align}
\phi(r,\theta,z,t)&=\phi(r,\theta,z)e^{-i\Omega t}, \nonumber \\
\Psi_i(r,\theta,z,t)&=\Psi_i(r,\theta,z)e^{-i\Omega t},
\end{align}
with $(i=r,\theta,z)$. The Laplacian of scalar and vector potentials are defined by
\begin{align}
\nabla^2\phi&=\nabla\cdot\nabla\phi, \nonumber \\
\nabla^2\vec{\Psi}&=\nabla(\nabla\cdot\vec{\Psi})-\nabla\times\nabla\times\vec{\Psi}.
\end{align}
The wave equations are given explicitly by
\begin{align}
\nabla^2\phi+\frac{\Omega^2}{v_l^2}\phi&=0, \nonumber \\
\nabla^2\Psi_r-\frac{1}{r^2}\Psi_r-\frac{2}{r^2}\frac{\partial\Psi_{\theta}}{\partial\theta}+\frac{\Omega^2}{v_t^2}\Psi_r&=0, \nonumber \\
\nabla^2\Psi_{\theta}-\frac{1}{r^2}\Psi_{\theta}+\frac{2}{r^2}\frac{\partial\Psi_r}{\partial\theta}+\frac{\Omega^2}{v_t^2}\Psi_{\theta}&=0, \nonumber \\
\nabla^2\Psi_z+\frac{\Omega^2}{v_t^2}\Psi_z&=0.
\end{align}
The equations for $\phi$ and $\Psi_z$ are separated, and for $\Psi_r$ and
$\Psi_{\theta}$ are coupled. The Laplacian $\nabla^2$ is given in cylindrical
coordinates. The solution for $\phi$ has the form
\begin{equation}
\phi(r,\theta,z,t)=R(r)\Theta(\theta)e^{i(qz-\Omega t)},
\end{equation}
where $R(r)$ and $\Theta(\theta)$ obey the equations
\begin{align}
\frac{\partial^2\Theta}{\partial\theta^2}+n^2\Theta&=0, \nonumber \\
\frac{\partial^2R}{\partial r^2}+\frac{1}{r}\frac{\partial R}{\partial r}+\left(\frac{\Omega^2}{v_l^2}-q^2\right)R-\frac{n^2}{r^2}R&=0,
\end{align}
with $(n=0,1,2,\cdots)$. The general solution has the form
\begin{equation}
\phi(r,\theta,z,t)=AJ_n(\eta_lr)\cos(n\theta)e^{i(qz-\Omega t)},
\end{equation}
with $\eta_l^2=\frac{\Omega^2}{v_l^2}-q^2$. The solution for
$\Theta$ can be either $\cos(n\theta)$ or $\sin(n\theta)$, and for $R(r)$ the
solution is the
Bessel functions of the first kind. Similar solution holds for
$\Psi_z$, where
\begin{equation}
\Psi_z(r,\theta,z,t)=BJ_n(\eta_rr)\cos(n\theta)e^{i(qz-\Omega t)},
\end{equation}
with $\eta_t^2=\frac{\Omega^2}{v_t^2}-q^2$.

The solutions for $\Psi_r$ and $\Psi_{\theta}$ have the form
\begin{align}
\Psi_r(r,\theta,z,t)&=\bar{\Psi}_r(r)\sin(n\theta)e^{i(qz-\Omega t)}, \nonumber \\
\Psi_{\theta}(r,\theta,z,t)&=\bar{\Psi}_{\theta}(r)\cos(n\theta)e^{i(qz-\Omega t)},
\end{align}
where
\begin{align}
\bar{\Psi}_r(r)&=DJ_{n-1}(\eta_tr)+CJ_{n+1}(\eta_tr), \nonumber \\
\bar{\Psi}_{\theta}(r)&=DJ_{n-1}(\eta_tr)-CJ_{n+1}(\eta_tr).
\end{align}
Here using $\cos(n\theta)$ in $\Psi_r$ implies $\sin(n\theta)$ in
$\Psi_{\theta}$, and vice versa.

We have four constants $A$, $B$, $C$ and $D$. We still free to add another
relation between the potentials. But the usually used relation of $\nabla\cdot\vec{\Psi}=0$ gives
a complex relation. Here we use the simple relation $\Psi_r=-\Psi_{\theta}$
that lead to
$D=0$.

The general solutions can be given by
\begin{align}
\phi(r,\theta,z,t)&=AJ_n(\eta_lr)\cos(n\theta)e^{i(qz-\Omega t)}, \nonumber \\
\Psi_r(r,\theta,z,t)&=CJ_{n+1}(\eta_tr)\sin(n\theta)e^{i(qz-\Omega t)}, \nonumber \\
\Psi_{\theta}(r,\theta,z,t)&=-CJ_{n+1}(\eta_tr)\cos(n\theta)e^{i(qz-\Omega t)}, \nonumber \\
\Psi_z(r,\theta,z,t)&=BJ_n(\eta_tr)\sin(n\theta)e^{i(qz-\Omega t)}.
\end{align}
The constants $A$, $B$, and $C$ are fixed from boundary conditions.

The displacement components are given in terms of the potentials by
\begin{align}
Q_r&=\frac{\partial\phi}{\partial
  r}+\frac{1}{r}\frac{\partial\Psi_z}{\partial\theta}-\frac{\partial\Psi_{\theta}}{\partial z}, \nonumber \\
Q_{\theta}&=\frac{1}{r}\frac{\partial\phi}{\partial\theta}-\frac{\partial\Psi_z}{\partial
  r}+\frac{\partial\Psi_r}{\partial z}, \nonumber \\
Q_z&=\frac{\partial\phi}{\partial
  z}-\frac{1}{r}\frac{\partial\Psi_r}{\partial\theta}+\frac{1}{r}\frac{\partial(r\Psi_{\theta})}{\partial
r}.
\end{align}
The boundary condition implies the surface, at $r=a$, to be free, which leads to the equation that relates $\Omega$, $q$ and $n$. Three cases can be treated, which are (i) torsional waves for $n=0$ where $Q_{\theta}$ is independent of $\theta$,
(ii) longitudinal waves for $n=0$ where $Q_r$ and $Q_z$ are independent of $\theta$, and (iii) flexural waves for $n=1$
where $Q_r$, $Q_{\theta}$ and $Q_z$ are dependent on $(r,\theta,z)$.

\subsection{Torsional Waves}

For axially symmetric torsional waves with $n=0$, which are the simplest case for elastic
waves in a rod, the displacement along the $\theta$ direction is defined by
\begin{equation}
Q_{\theta}(r,z,t)=-\frac{\partial}{\partial r}\Psi_z,
\end{equation}
and $Q_{r}=Q_{z}=0$, where the potentials $\phi$, $\Psi_r$ and $\Psi_{\theta}$ are taken to be
zero. Here the solution for the potential is $\Psi_z=\Psi_z(r,z,t)$, and has the form
\begin{equation}
\Psi_z=BJ_0(\eta_tr)e^{i(qz-\Omega t)},
\end{equation}
where $B$ is an amplitude, and $\eta^2_t=\frac{\Omega^2}{v_t^2}-q^2$. The displacement reads
\begin{equation}
Q_{\theta}(r,z,t)=B\eta_tJ_1(\eta_tr)e^{i(qz-\Omega t)}.
\end{equation}

The force-free surface leads to the condition \cite{Achenbach1975} $\eta_t aJ_2(\eta_t a)=0$, where
either $\eta_t a=0$ or $J_2(\eta_ta)=0$. The case of $\eta_t a=0$ gives the solution
\begin{equation}
\Psi_z\approx-B\frac{r^2\eta_t^2}{6}e^{i(q_0z-\Omega t)},
\end{equation}
with $q_0=\Omega/v_t$. Hence, we have
\begin{equation}
Q_{\theta}^0\approx\frac{1}{3}B\eta_t^2re^{i(q_0z-\Omega t)}.
\end{equation}
The condition $J_2(\eta_t a)=0$ gives a set of roots, e.g.,
$\varepsilon_1=5.136$, $\varepsilon_2=8.417$, $\varepsilon_1=11.620$,
$\cdots$, and we write $\varepsilon_n=\eta_ta$, where we can write the solutions
as
\begin{equation}
Q_{\theta}^n=B\frac{\varepsilon_n}{a}J_1(\varepsilon_n r/a)e^{i(q_nz-\Omega t)},
\end{equation}
with $q^2=\frac{\Omega^2}{v_t^2}-\frac{\varepsilon_n^2}{a^2}$.

\subsection{Longitudinal Modes}

For longitudinal acoustic modes with $n=0$ from symmetry consideration we have
$Q_r=Q_r(r,z,t)$, $Q_z=Q_z(r,z,t)$ and $Q_{\theta}=0$. Here we have $\Psi_r=\Psi_z=0$,
and hence the displacements are related to the potentials by
\begin{align}
Q_r&=\frac{\partial\phi}{\partial r}-\frac{\partial\Psi_{\theta}}{\partial
  z}, \nonumber \\
Q_z&=\frac{\partial\phi}{\partial z}+\frac{1}{r}\frac{\partial(r\Psi_{\theta})}{\partial
    r}.
\end{align}
The solutions for the potentials are
\begin{align}
\phi=AJ_0(\eta_l r)e^{i(qz-\Omega t)}, \nonumber \\
\Psi_{\theta}=-CJ_1(\eta_t r)e^{i(qz-\Omega t)},
\end{align}
where $\eta_i^2=\frac{\Omega^2}{v_i^2}-q^2$, for $(i=l,t)$. The requirement of the stress to vanish on the fiber boundary, $(r=a)$, lead
to the Pochhammer frequency equation
\begin{equation}\label{Long}
(q^2-\eta_t^2)^2\frac{\eta_laJ_0(\eta_la)}{J_1(\eta_la)}+4q^2\eta_l^2\frac{\eta_taJ_0(\eta_ta)}{J_1(\eta_ta)}=2\eta_l^2(q^2+\eta_t^2).
\end{equation}
In the limit of $qa\ll 1$ one gets the sound wave linear dispersion
$\Omega\approx v_tq$. The displacements are given explicitly by
\begin{align}
Q_r&=-\left\{A\eta_lJ_1(\eta_l
r)-iqCJ_1(\eta_tr)\right\}e^{i(qz-\Omega t)}, \nonumber \\
Q_{\theta}&=0, \nonumber \\
Q_z&=\left\{iqAJ_0(\eta_l r)-\eta_tCJ_0(\eta_tr)\right\}e^{i(qz-\Omega t)}.
\end{align}
From boundary condition we can get also relations between $A$ and $C$. Namely,
the vanishing of the stress and the strain on the surface yields \cite{Achenbach1975}
\begin{equation}\label{Relations}
2iq\eta_lJ_1(\eta_la)A=\left[\eta_t^2-q^2\right]J_1(\eta_ta)C.
\end{equation}


\end{document}